\documentclass[fleqn,usenatbib]{mnras}
\usepackage{newtxtext,newtxmath}
\usepackage{amsmath}
\usepackage{graphics,graphicx}
\usepackage{natbib}
\usepackage{setspace}
\usepackage{caption,setspace}
\newcommand{\angstrom}{\mbox{\normalfont\AA}}
\usepackage[dvipsnames]{xcolor}
\usepackage{caption}
\usepackage{subcaption}
\usepackage{mathtools}
\usepackage{enumitem}
\usepackage{lipsum}
\setlist[itemize]{leftmargin=*}

\usepackage{floatrow}
\usepackage[outercaption]{sidecap}    
\sidecaptionvpos{figure}{t}

\usepackage[T1]{fontenc}
\usepackage{ae,aecompl}
\pdfoutput=1

\title[The impact of AGN-driven winds on galaxy sizes]{The impact of AGN-driven winds on physical and observable galaxy sizes}
\author[R.K. Cochrane et al.]{R. K. Cochrane,$^{1}$\thanks{E-mail:rcochrane@flatironinstitute.org}
D. Angl\'es-Alc\'azar,$^{2,1}$
J. Mercedes-Feliz,$^{2}$
C. C. Hayward,$^{1}$
C.-A. Faucher-Gigu\`ere,$^{3}$
\newauthor
S. Wellons,$^{4}$
B. A. Terrazas,$^{5}$ 
A. Wetzel,$^{6}$ 
P. F. Hopkins,$^{7}$ 
J. Moreno,$^{8}$ 
K.-Y. Su,$^{9}$ 
and R. S. Somerville$^{1}$
\vspace{0.2cm}\\
$^{1}$Center for Computational Astrophysics, Flatiron Institute, 162 Fifth Avenue, New York, NY 10010, USA\\
$^{2}$Department of Physics, University of Connecticut, 196 Auditorium Road, U-3046, Storrs, CT 06269-3046, USA\\
$^{3}$Department of Physics and Astronomy and CIERA, Northwestern University, Evanston, IL 60208, USA\\
$^{4}$Department of Astronomy, Van Vleck Observatory, Wesleyan University, 96 Foss Hill Drive, Middletown, CT 06459, USA\\
$^{5}$Columbia Astrophysics Laboratory, Columbia University, 550 West 120th Street, New York, NY 10027, USA\\
$^{6}$Department of Physics \& Astronomy, University of California, Davis, CA 95616, USA\\
$^{7}$TAPIR, MC 350-17, California Institute of Technology, Pasadena, CA 91125, USA\\
$^{8}$Department of Physics and Astronomy, Pomona College, Claremont, CA 91711, USA\\
$^{9}$Black Hole Initiative, Harvard University, 20 Garden St., Cambridge, MA 02138, USA}
\date{Accepted XXX. Received YYY; in original form ZZZ}
\pubyear{2023}
\begin{document}
\label{firstpage}
\pagerange{\pageref{firstpage}--\pageref{lastpage}}
\maketitle
\begin{abstract}
Without AGN feedback, simulated massive, star-forming galaxies become too compact relative to observed galaxies at $z\lesssim2$. In this paper, we perform high-resolution re-simulations of a massive ($M_{\star}\sim10^{11}\,\rm{M_{\odot}}$) galaxy at $z\sim2.3$, drawn from the Feedback in Realistic Environments (FIRE) project. In the simulation without AGN feedback, the galaxy experiences a rapid starburst and shrinking of its half-mass radius. We experiment with driving mechanical AGN winds, using a state-of-the-art hyper-Lagrangian refinement technique to increase particle resolution. These winds reduce the gas surface density in the inner regions of the galaxy, suppressing the compact starburst and maintaining an approximately constant half-mass radius. Using radiative transfer, we study the impact of AGN feedback on the magnitude and extent of the multi-wavelength continuum emission. When AGN winds are included, the suppression of the compact, dusty starburst results in lowered flux at FIR wavelengths (due to decreased star formation) but increased flux at optical-to-near-IR wavelengths (due to decreased dust attenuation, in spite of the lowered star formation rate), relative to the case without AGN winds. The FIR half-light radius decreases from $\sim1\,\rm{kpc}$ to $\sim0.1\,\rm{kpc}$ in $\lesssim40\,\rm{Myr}$ when AGN winds are not included, but increases to $\sim2\,\rm{kpc}$ when they are. Interestingly, the half-light radius at optical-NIR wavelengths remains approximately constant over $35\,\rm{Myr}$, for simulations with and without AGN winds. In the case without winds, this occurs despite the rapid compaction, and is due to heavy dust obscuration in the inner regions of the galaxy. This work highlights the importance of forward-modelling when comparing simulated and observed galaxy populations.
\end{abstract}
\begin{keywords}
galaxies: evolution -- galaxies: active -- quasars: supermassive black holes -- ISM: jets and outflows -- radiative transfer 
\end{keywords}
\section{Introduction}
Decades of work have contributed to our current understanding of the co-evolution of supermassive black holes (SMBHs) and their host galaxies (see \citealt{Heckman2014} and \citealt{DiMatteo2023} for reviews). On the scales of individual galaxies, black hole mass correlates with the stellar velocity dispersion (the `$M_\mathrm{BH}-\sigma$' relation; \citealt{Ferrarese2000,Gebhardt2000,Ferrarese2001,Tremaine2002,Gultekin2009}) as well as with the mass of the stellar bulge \citep{Kormendy1995,Magorrian1998,McLure2002,Marconi2003,Haring2004}. These two key scaling relations establish that black holes and galaxies build up their mass together. This is supported by measurements of the cosmic evolution of the volume-averaged star formation rate density (SFRD) and the SMBH accretion rate density (BHARD, as inferred from integrating AGN luminosity functions, using Soltan's argument; \citealt{Soltan1982}, e.g. \citealt{Merloni2008,Shankar2009, Delvecchio2014}). The SFRD and BHARD display remarkably similar forms \citep{Boyle1998,Shankar2009,Madau2014}: both increase back to $z=2$ and flatten and decrease thereafter, broadly following the availability of cold gas. \\
\indent Feedback from active galactic nuclei (AGN) is believed to play a role in this co-evolution. `Jet-mode' feedback is seen directly via the strong synchotron jets observed at radio wavelengths \citep{Best2012}, and via X-ray bubbles and cavities shock-heated by these jets (see the review by \citealt{Fabian2012}). This mode of feedback, found in the central galaxies of groups and clusters, is often referred to broadly as `preventive', as it heats gas and therefore prevents efficient cooling and star formation \citep[see the review by][]{McNamara2012}. `Radiative-mode' feedback, signposted by observations of outflows of molecular, neutral atomic and ionised gas (see e.g. \citealt{Sturm2011, Rupke2011,Cicone2014,Zakamska2014,Fiore2017}), is thought to play some role in removing material from galaxies. However, many details about this feedback and its coupling to the host galaxy's ISM and beyond (e.g. how quickly energy is radiated away) are not fully understood \citep{Harrison2018}. One difficulty in modelling feedback ab-initio is the vast dynamic range to be spanned, from the sub-parsec scale accretion disk of the black hole to the intergalactic medium (IGM) or intracluster medium (ICM). While most galaxy formation simulations require some form of AGN feedback to reproduce empirical results such as the high mass end of the stellar mass function and fractions of quiescent galaxies \citep[e.g.][]{Springel2005a,Springel2005b,Hirschmann2014,Vogelsberger2014,Schaye2015,Dubois2016,Weinberger2017,Weinberger2018,Pillepich2018a,Dave2019}, implementations rely on calibrated but not always physically-motivated `sub-grid' modelling. \\
\indent AGN-driven outflows and their observational signatures have been modelled analytically and in idealized simulations \citep[e.g.][]{Ciotti1997,Ciotti2001,Faucher-Giguere2012,Costa2014,Nims2015,Hopkins2016,Richings2018}. Ejective feedback can increase galaxy sizes since following the removal of gas, galaxy mergers tend to be `dry' (gas-poor) and likely to form extended stellar envelopes \citep{Choi2018}. Fast winds can also act to `puff up’ the central regions of galaxies \citep{Choi2018}. Hence, massive simulated galaxies with AGN have a lower fraction of stellar mass formed in-situ, with flatter central stellar mass densities \citep{Dubois2016}. As a result, the choice of AGN feedback implementation can impact the positions of simulated galaxies on the stellar mass-galaxy size relation. Cosmological simulations are therefore often calibrated to recover realistic low-redshift galaxy sizes, in addition to the $z\approx0$ galaxy stellar mass function \citep[e.g.][]{Crain2015,Pillepich2018a}. \\
\indent In parallel to these strides in simulations, observational studies have constrained distributions of galaxy sizes out to high redshifts \citep[e.g.][]{Bouwens2004,Ferguson2004,Trujillo2006,Trujillo2007,Buitrago2008,VanDokkum2008,VanDokkum2010,VanDerWel2014,Mowla2019,Whitney2021,Barone2022,Hamadouche2022}. The evolution of the sizes of disk galaxies between intermediate and low redshift is thought to be driven by accretion of cold gas and subsequent star formation, while the (more substantial) evolution in the sizes of early-type galaxies is now believed to be driven largely by major and minor mergers \citep{Khochfar2006,Naab2010,Bluck2012,Cimatti2012,Whitney2021}, with progenitor effects also contributing to a perceived evolution \citep{vanDokkum2001,Carollo2013a,Ji2022}. As discussed above, AGN feedback may also play a role in the size growth of galaxies, at least in certain evolutionary phases \citep[see also][]{Fan2008,VanderVlugt2019}.\\
\indent The Feedback in Realistic Environments (FIRE\footnote{\url{https://fire.northwestern.edu}}) project incorporate stellar feedback based on stellar synthesis models without the need to fine-tune parameters \citep{Hopkins2014,Hopkins2017,Hopkins2022}. Stellar feedback processes implemented include core collapse and Type Ia supernovae, winds from young (O/B) and evolved (AGB) stars, and radiation (photo-ionization, photoelectric heating, and radiation pressure). 
\cite{Parsotan2021} studied the sizes and central surface densities of four massive FIRE galaxies between $z\sim2.75$ and $z\sim1.25$ (these galaxies were simulated by \citealt{Angles-Alcazar2017} and reach $M_{\star}\sim10^{11}\,\rm{M_{\odot}}$ by $z=2$). For the most robust comparison with observations, they performed radiative transfer on the simulated galaxies, and derived these sizes and surface densities using the same methods as used for observed galaxies (following \citealt{VanDerWel2014}). In general, the simulated galaxies were found to lie below empirically-derived size-mass and size-density relations, with particular deviations (of an order of magnitude or more) towards the lowest redshifts/highest stellar masses studied. Less massive simulated FIRE galaxies \citep[e.g.][]{Sanderson2020,Bellardini2021} do not show such inconsistencies with observed galaxies \citep{El-Badry2016,Rohr2022}. \cite{Parsotan2021} commented that these results suggest that AGN feedback might help ease tensions for the most massive galaxies (see also \citealt{Wellons2019}). At the time of that work, AGN feedback had not been implemented within FIRE. \\
\indent The first attempt at studying AGN feedback in a system modelled with FIRE physics was presented by \cite{Hopkins2016}. They studied the interaction of AGN feedback and the multi-phase ISM of a gas-rich nuclear disk (on scales of $\sim0.1-100\,\rm{pc}$). The implemented winds generated a polar cavity, eventually evacuating the nuclear region and suppressing nuclear star formation and black hole growth. Making progress toward a fully-consistent implementation of `quasar-mode' feedback within FIRE on galaxy scales, \cite{Torrey2020} modelled the coupling of feedback from fast nuclear winds to the ISM of an idealised simulated galaxy, including both a kinetic component from the winds and Compton heating/cooling. In their model, particles are spawned symmetrically $0.1\,\rm{pc}$ from the central black hole, with velocity vectors pointing outwards. Each spawned wind particle has temperature $T=10^{4}\,\rm{K}$, solar metallicity and velocity $v=0.1c$. Compared to adding mass, momentum and energy from the wind to existing particles, spawning new particles has the advantage of enabling better resolution of the wind shock. Even when the density of gas around the black hole is low, the feedback is injected locally. Importantly, the wind modelling was implemented within a realistic galaxy disk drawn from the FIRE suite, enabling detailed studies of how the wind couples to the gas, taking into account its geometry and porosity. \cite{Torrey2020} showed that the ambient ISM is shocked to high temperatures by the wind, creating a low-density cavity at the galaxy's center, following expectations from analytical work \citep[e.g.][]{Faucher-Giguere2012} and one-dimensional hydrodynamical simulations \citep{Ciotti2001}. For the most massive black holes accreting near to the Eddington limit, this ISM cavity can span $1-10\,\rm{kpc}$ over tens of Myr. When implemented in cosmological simulations, this cavity-opening AGN wind model may prevent the formation of overly-dense stellar cores such as those seen in the massive FIRE galaxies studied in previous papers \citep[e.g.][]{Angles-Alcazar2017,Wellons2019,Parsotan2021,Byrne2023}. Very recent work has focused on exploring the parameter space of AGN feedback capable of quenching FIRE-simulated galaxies across different halo mass scales \citep{Su2020,Su2021,Wellons2022}.\\
\indent This paper forms part of a series presenting our work implementing a wind model similar to that presented by \cite{Torrey2020} in the cosmological context of FIRE galaxies. \cite{Angles-Alcazar} describe the technical details of the hyper-refined accretion-driven wind model, which is set within a multi-phase ISM including feedback from supernovae, stellar winds and radiation. The new implementation builds upon the foundations set by \cite{Torrey2020}, now capturing the propagation and impact of the winds from the scales of the inner nuclear region ($<10\,\rm{pc}$) to the circumgalactic medium. A single galaxy drawn from the FIRE simulations was used for this study (the central galaxy of halo A4; \citealt{Angles-Alcazar2017, Feldmann2017}). This galaxy forms considerable stellar mass early in its evolution, reaching $M_{\star}\sim10^{10.8}\,\rm{M_{\odot}}$ at $z = 2.28$. \cite{Parsotan2021} showed that the half-mass radius of the galaxy decreases from $\sim2\,\rm{kpc}$ at $z=2.75$ (where it lies on the size-stellar mass relation derived using large samples of observed galaxies) to $\lesssim1\,\rm{kpc}$ at $z=2.25$. \cite{Angles-Alcazar} chose to implement AGN-driven winds at the time when the galaxy undergoes its strongest starburst phase and rapid sub-pc scale accretion onto the central black hole \citep{Angles-Alcazar2021}, to investigate whether winds are capable of suppressing the rapid compaction process. In other papers in this series, we study the possibility of these winds inducing positive feedback \citep{Mercedes-Feliz2022a}, the formation of dense stellar clumps \citep{Mercedes-Feliz2022}, and the impact of quasar winds on $\rm{H}\alpha$ emission \citep{Terrazas2023}. These ultra-high resolution simulations focusing on a single quasar phase complement recent FIRE cosmological simulations down to $z=0$ exploring the impact of a variety of black hole accretion and feedback implementations across halo mass \citep{Wellons2022}. \\
\indent In this paper, we compare the physical (half-mass) and observable (half-light) sizes of the simulated galaxy with and without constantly-driven AGN winds, across $\sim35\,\rm{Myr}$. In Section \ref{sec:FIRE_sims}, we provide an overview of the FIRE simulations and describe the new implementation of quasar-driven winds. We study the impact of different wind models on the simulated galaxy in terms of its modelled stellar mass, SFR and gas mass, as well as their spatial extents, in Section \ref{sec:FIRE_physical_properties}. In Section \ref{sec:results}, we describe the methods used to generate synthetic images of the simulated galaxies. We compare emission maps, radial profiles and half-light sizes of the galaxy simulated with and without quasar-driven winds. In Section \ref{sec:conclusions}, we summarise our results and draw conclusions.\\
\indent Throughout this work, we assume a $\Lambda\rm{CDM}$ cosmology with $H_{0}=69.7\,\rm{km\,s^{-1}\,Mpc^{-1}}$, $\Omega_{\rm{M}}=1-\Omega_{\Lambda}=0.2821$, $\Omega_{\rm{b}}=0.0461$, $\sigma_{8}$, and $n_{\rm{s}}=0.9646$ \citep{Hinshaw2013}.

\begin{figure*}
\centering
\includegraphics[width=4.45cm]{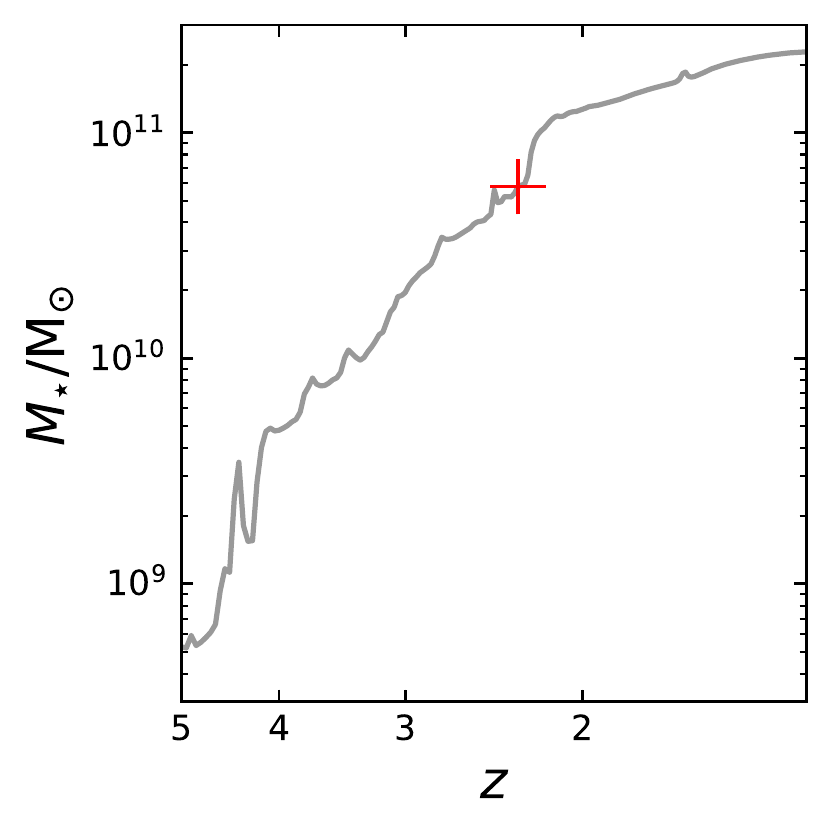}
\hspace{-0.2cm}
\includegraphics[width=4.45cm]{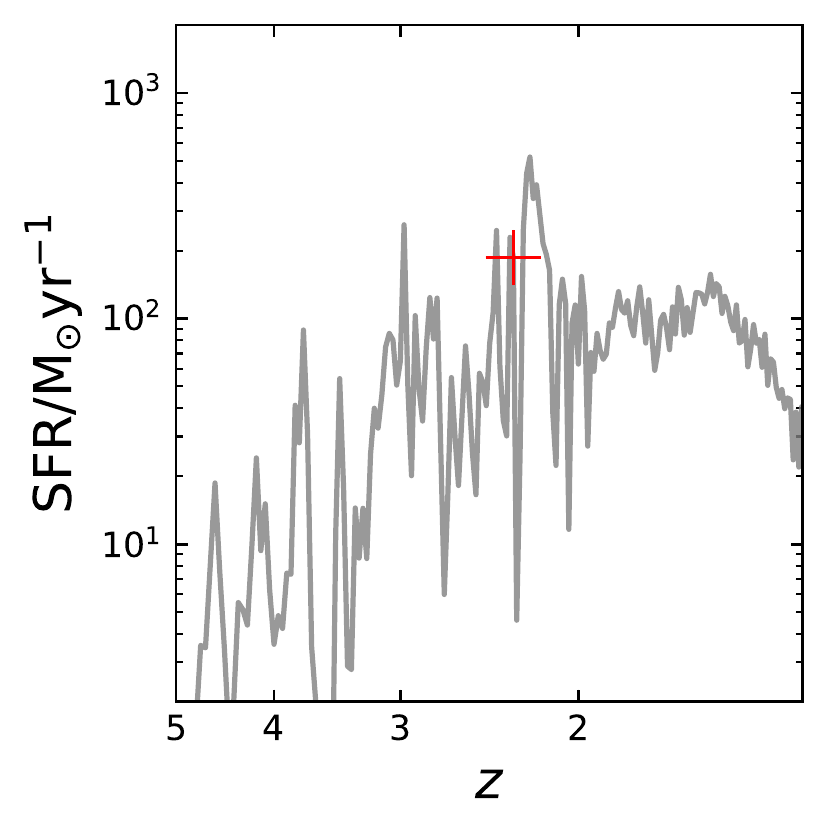}
\hspace{-0.2cm}
\includegraphics[width=4.45cm]{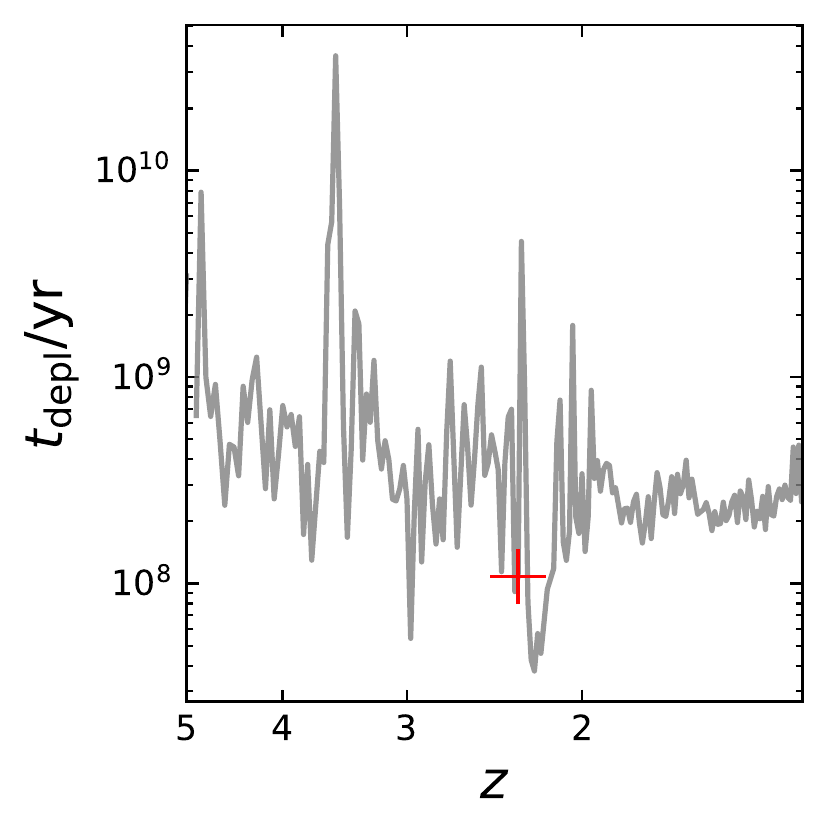}
\hspace{-0.2cm}
\includegraphics[width=4.45cm]{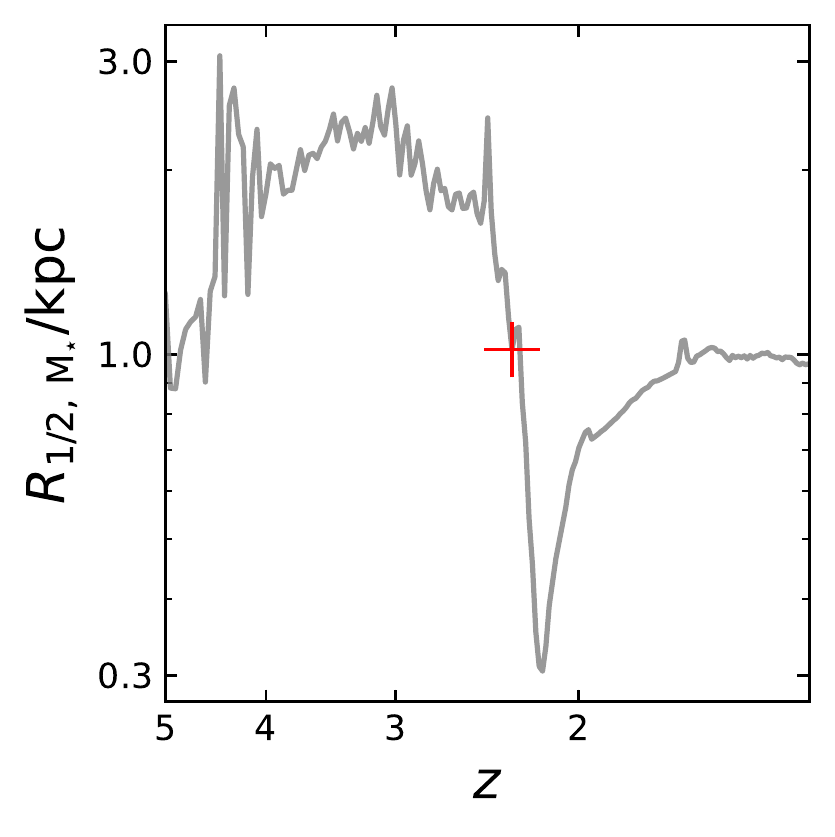}
\hspace{-0.2cm}
\caption{The evolution of the stellar mass, SFR, gas depletion time and half-mass radius of our simulated galaxy from $z=5$ to $z=1.2$ within the original zoom-in simulation from which we draw our initial conditions. All quantities are calculated within a 3D aperture of radius $0.1\,R_{\rm{vir}}$. The red cross highlights the redshift chosen to start our re-simulation including hyper-refined AGN-driven winds ($z=2.32$). In the original simulation, the galaxy experiences a global maximum in SFR and a minimum in gas depletion timescale within the following $\sim100\,\rm{Myr}$, with an accompanying rapid decrease in half-mass radius. In this paper, we explore the impact of turning on AGN winds just before this starburst event.}
\label{fig:fiducial_fig}
\end{figure*}

\section{FIRE simulations of a massive galaxy at z=2.3}\label{sec:FIRE_sims}
In this paper, we study the impact of AGN-driven winds on the host galaxy using a novel implementation described fully in \cite{Angles-Alcazar}. We focus on a short period of time ($\sim40\,\rm{Myr}$, beginning at $z=2.32$) in the evolution of one of the massive galaxies
that were originally simulated by \cite{Feldmann2017a} using the FIRE-1 model and re-simulated by \cite{Angles-Alcazar2017} using the FIRE-2 model (in other papers, this galaxy is labeled A4). The evolution of the physical properties of this galaxy is shown in Figure \ref{fig:fiducial_fig} (see also \citealt{Cochrane2022} and the detailed analysis of the structural and dynamical properties of the galaxy presented by \citealt{Wellons2019}). At $z=2.319$, the stellar, gas and halo masses of the galaxy are: $\log_{10}(M_{\star}/\rm{M_{\odot}})=10.76$, $\log_{10}(M_{\rm{gas}}/\rm{M_{\odot}})=10.30$, and $\log_{10}(M_{\rm{halo}}/\rm{M_{\odot}})=12.29$. Within the subsequent $\sim100\,\rm{Myr}$, the galaxy experiences its strongest starburst event. As shown in Figure \ref{fig:fiducial_fig}, at $z=2.233$ the star formation rate reaches $520\,\rm{M_{\odot}\,yr^{-1}}$, the highest star formation rate in its simulated evolution, and a corresponding minimum in gas depletion time of $38\,\rm{Myr}$. The half-mass radius decreases from $2\,\rm{kpc}$ to $\sim0.3\,\rm{kpc}$, accompanied by an increase in stellar circular velocity (from $\sim300\,\rm{km\,s^{-1}}$ to $\sim600\,\rm{km\,s^{-1}}$ at $1\,\rm{kpc}$), and a corresponding increase in stellar velocity dispersion (see \citealt{Angles-Alcazar} for further discussion).

\subsection{Fiducial simulation without AGN feedback}\label{sec:fiducial_model}
The FIRE project is a suite of state-of-the-art hydrodynamical cosmological zoom-in simulations described fully in \citet[][FIRE-1]{Hopkins2014}, \citet[][FIRE-2]{Hopkins2017} and \citet[][FIRE-3]{Hopkins2022}. The fiducial simulations without winds, described fully by \cite{Angles-Alcazar2017}, use the FIRE-2 physics model \citep{Hopkins2017}. FIRE-2 uses the ``meshless finite mass'' mode of the $N$-body+hydrodynamics code GIZMO\footnote{\url{http://www.tapir.caltech.edu/~phopkins/Site/GIZMO.html}} \citep{hopkins2015}; gravitational forces are computed following the methods presented in \cite{Hopkins2013}, using an 
improved version of the parallel TreeSPH code GADGET-3 \citep{Springel2005}. Cooling and heating processes including  free-free, photoionization/recombination, Compton, photo-electric, metal-line, molecular and fine-structure processes are modelled from $T=10\,\rm{K}$ to $T=10^{10}\,\rm{K}$. Star particles form from locally self-gravitating, molecular, Jeans unstable gas above a minimum hydrogen number density $n_{H}\geq1000\,\rm{cm}^{-3}$. Each star particle represents a single stellar population with known mass, age, and metallicity, injecting feedback locally in the form of mass, momentum, energy, and metals from Type Ia and Type II Supernovae (SNe), stellar winds, photoionization and photoelectric heating, and radiation pressure, with all feedback quantities and their time dependence taken directly from the {\sc{starburst}99} population synthesis model \citep{Leitherer1999}. \\
\indent The baryonic mass resolution is $3.3\times10^{4}\,\rm{M}_{\odot}$. Softenings for gas (minimum adaptive force softening), stellar and dark matter particles are given by: $\epsilon_{\rm{gas}}^{\rm{min}}=0.7\,\rm{pc}$, $\epsilon_{\star}=7\,\rm{pc}$, and $\epsilon_{\rm{DM}}=57\,\rm{pc}$.

\begin{table}
\begin{center}
\begin{tabular}{c|c|c|c|c|c}
\hline
\hline
Name & $\lambda_{\rm{Edd}}$ & $\epsilon_{k}$ & $\eta_{k}$ & $\dot{M}_{\rm{w}}/\rm{M_{\odot}\,yr^{-1}}$ \\ 
\hline
\bf{noAGN} & - & - & - & - \\
m0.1e0.5 & $1$ & $0.005$ & $0.1$ & $2.2$ \\
m1e5 & $1$ & $0.05$ & $1$ & $22.2$ \\
\bf{m2e10} & $1$ & $0.1$ & $2$ & $44.4$  \\
m4e20 & $1$ & $0.2$ & $4$ & $88.8$ \\
m10e50 & $1$ & $0.5$ & $10$ & $222$ \\
\hline
\hline
\end{tabular}
\caption{Parameters used for the seven simulations, which are described fully by \protect\cite{Angles-Alcazar}. The fiducial model, named {\bf{noAGN}}, models black hole accretion but not feedback. The six implementations of AGN-driven winds represent different choices of the mass outflow rate, $\dot{M}_{\rm{w}}$. Simulations are are named according to two parameters, $\eta_{\rm{k}}$ and $\epsilon_{k}$. $\lambda_{\rm{Edd}}=\dot{M}_{\rm{BH}}/\dot{M}_{\rm{Edd}}$ is the black hole accretion rate, in units of the Eddington rate, and $\epsilon_{k}=\dot{E}_{\rm{w}}/L_{\rm{bol}}$ is the kinetic feedback efficiency. $\eta_{k}=\dot{M}_{\rm{W}}/\dot{M}_{\rm{BH}}$ is the mass loading factor. In this paper, we primarily study model {\bf{m2e10}}.}
\label{Table:sims_params}
\end{center}
\end{table}

\subsubsection{Modelling an accreting black hole} 
The FIRE-2 simulations with BHs introduced in \cite{Angles-Alcazar2017} modelled supermassive black hole (SMBH) growth following a gravitational torque-accretion model \citep{Hopkins2011, Angles-Alcazar2013,Angl2018} and did not include AGN feedback (see also \citealt{Catmabacak2022,Byrne2023}). For simplicity, in the numerical experiments presented here we assume a constant accretion rate. In both fiducial and wind simulations, the black hole is modelled as a collisionless particle with initial mass $M_{\rm{BH}}=10^{9}\,\rm{M_{\odot}}$. The black hole accretion rate, $\dot{M}_{\rm{BH}}$, is constant throughout the modelled $\sim40\,\rm{Myr}$ quasar phase at the Eddington rate ($\lambda_{\rm{Edd}}=1$) and defined as follows:
\begin{equation}   \dot{M}_{\rm{BH}}=\lambda_{\rm{Edd}}\times\dot{M}_{\rm{Edd}} = \lambda_{\rm{Edd}}\times\frac{4\pi\,G\,m_{\rm{p}}\,M_{\rm{BH}}}{\epsilon_{\rm{r}}\,\sigma_{\rm{T}}\,c},
\end{equation}
where $G$ is the gravitational constant, $m_{\rm{p}}$ is the proton mass, $\epsilon_{\rm{r}}=0.1$ sets the radiative efficiency, $\sigma_{\rm{T}}$ is the Thomson scattering cross-section, and $c$ is the speed of light.

\begin{figure*} 
\centering
\includegraphics[width=0.95\columnwidth]{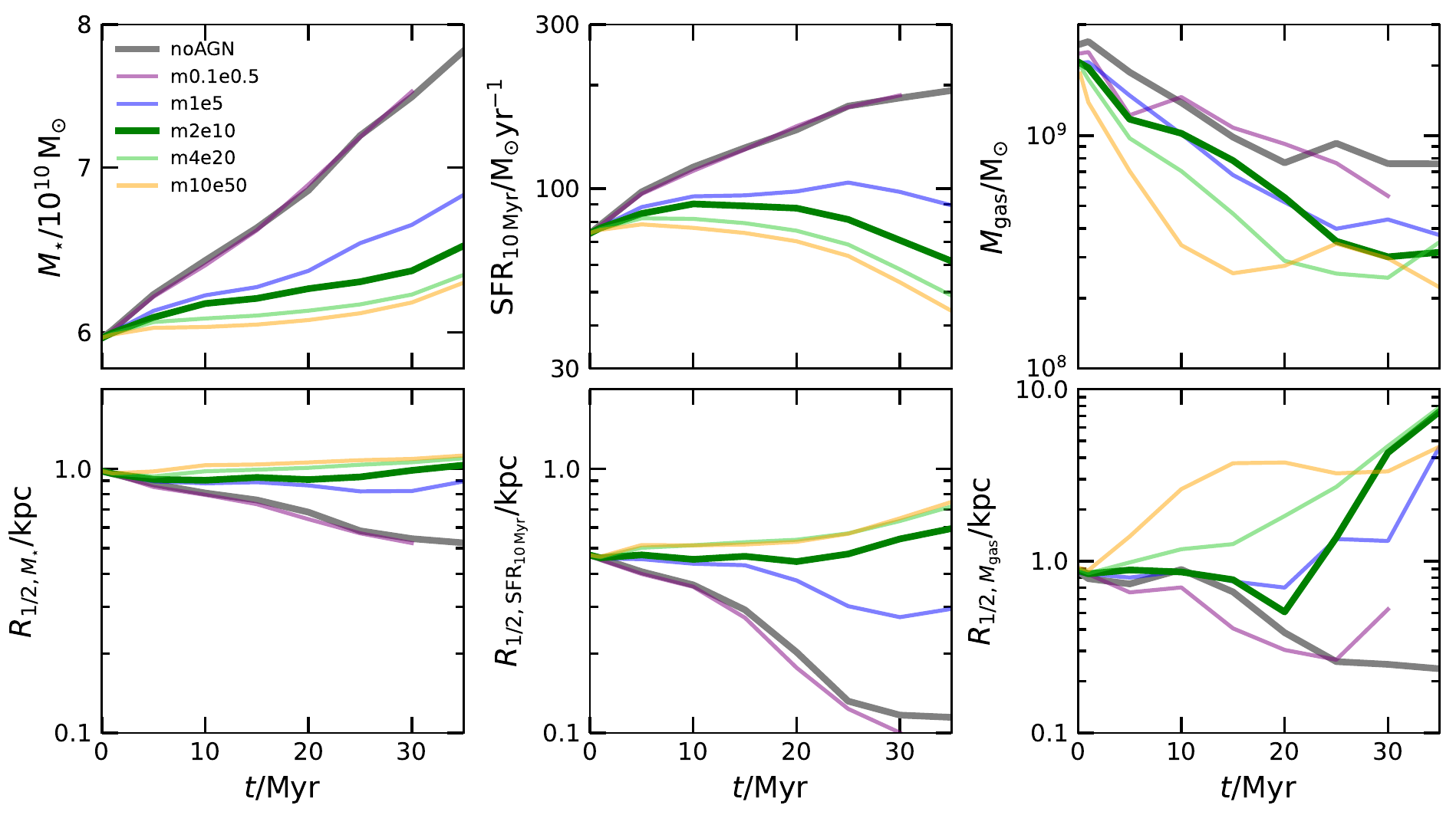}
\caption{Top: the evolution of stellar mass, SFR (averaged over $10\,\rm{Myr}$), and gas mass with time, for the fiducial simulation (thick grey) and five feedback implementations listed in Table \ref{Table:sims_params}.
Bottom: time evolution of radii containing half the stellar mass, SFR, and gas mass. In the fiducial simulation without AGN-driven winds, the galaxy undergoes a rapid decrease in half-mass radius, shrinking from $\sim1\,\rm{kpc}$ to $\sim0.5\,\rm{kpc}$ in just $\sim30\,\rm{Myr}$. AGN-driven winds with kinetic feedback efficiency $\epsilon_{k} \gtrsim 5\%$ are able to prevent the decrease in size seen in the no-feedback simulation, maintaining the half-mass radius at $0.8-1\,\rm{kpc}$. In the rest of this paper, we focus our analysis on simulation {\bf{m2e10}}, shown by the thick dark green line.}
\label{fig:daniel_re_vs_t}
\end{figure*}

\subsection{Modelling quasar-driven winds}
\label{sec:FIRE_sims_AGN_winds}
In our simulations with and without AGN winds, the FIRE-2 physics, black hole accretion model, particle resolution and softening lengths described in Section \ref{sec:fiducial_model} are implemented. The key difference for the AGN wind simulations is the addition of quasar-driven winds, described here. \\
\indent Winds are modelled using a particle spawning technique, such that feedback is injected locally and isotropically, and the wind-ISM interaction is captured regardless of gas geometry \citep{Torrey2020,Su2021,Angles-Alcazar}. The different simulations model winds with different choices of mass outflow rate, $\dot{M}_{\rm{W}}$. Other parameters characterising the wind are the initial velocity, $v_{\rm{W}}$, the post-shock velocity, $v_{\rm{sh}}$, post-shock temperature, $T_{\rm{sh}}$, and its geometry. The mass outflow rate is given by \citep[e.g.][]{Ostriker2010,Choi2012}
\begin{equation}
    \dot{M}_{\rm{W}}\equiv\frac{1-f_{\rm{acc}}}{f_{\rm{acc}}} \times\dot{M}_{\rm{BH}},
\end{equation}
where $f_{\rm{acc}}$ represents the fraction of inflowing gas that is accreted (rather than ejected) by the black hole. All other parameters follow from this choice. The mass loading factor, $\eta_{\rm{k}}$, is defined as
\begin{equation}
    \eta_{\rm{k}} \equiv \frac{\dot{M}_{\rm{W}}}{\dot{M}_{\rm{BH}}}=\frac{1-f_{\rm{acc}}}{f_{\rm{acc}}}.
\end{equation}
The momentum loading factor, $p_{\rm{k}}$, is given by
\begin{equation}
    p_{\rm{k}} \equiv \frac{\dot{P}_{\rm{W}}}{L_{\rm{bol}}/c} = \frac{v_{\rm{w}}}{\epsilon_{\rm{r}}c}\Bigg(\frac{1-f_{\rm{acc}}}{f_{\rm{acc}}}\Bigg).
\end{equation}
The energy loading factor, $\epsilon_{\rm{k}}$, is given by
\begin{equation}
    \epsilon_{\rm{k}} \equiv \frac{\dot{E}_{\rm{W}}}{L_{\rm{bol}}} = \frac{1}{2\epsilon_{r}}\Big(\frac{v_{\rm{w}}}{c}\Big)^{2}\Bigg(\frac{1-f_{\rm{acc}}}{f_{\rm{acc}}}\Bigg).
\end{equation}
Here, the momentum flux, $\dot{P}_{\rm{W}}$, is given by $\dot{P}_{\rm{W}}\equiv\dot{M}_{\rm{W}}\,v_{\rm{w}}$ and the kinetic energy of the winds, $\dot{E}_{\rm{W}}$, is given by $\dot{E}_{\rm{W}}\equiv\dot{M}_{\rm{W}}\,v_{\rm{w}}^{2}/2$. The different simulations presented in Table \ref{Table:sims_params} model different outflow rates, with correspondingly different momentum and energy loading factors. \\
\indent In all our wind model variations, the wind is launched with initial velocity $v_{\rm{W}}=30,000\,\rm{km\,s^{-1}}$ and temperature $T_{\rm{W}}\sim10^{4}\,\rm{K}$. Upon interaction with the ambient medium, a strong shock develops. The post-shock gas velocity is $v_{\rm{sh}}=v_{\rm{w}}/4=7500\,\rm{km\,s^{-1}}$, and the post-shock gas temperature is $T_{\rm{sh}}\sim1.2\times10^{10}\,\rm{K}$ (both calculated assuming Rankine-Hugoniot jump conditions for monatomic gas; \citealt{Faucher-Giguere2012}). The wind particle mass, $m_{\rm{w}}$, is 20 times lower than that of the gas particles: $m_{\rm{w}}\sim1000\,\rm{M_{\odot}}\,h^{-1}$ (i.e. the wind is more highly-resolved). The particles interact with the ISM immediately upon injection. Once the wind particles have transferred most of their energy and momentum to the gas (defined as when $v_{\rm{w}}$ falls to $<10\%$ of its initial value, i.e. when $v_{\rm{w}}<750\,\rm{km\,s^{-1}}$), they merge onto their nearest regular gas particle. This serves to minimise the computational cost of the simulation. All conserved quantities are maintained in this process.\\
\indent For both simulations with and without AGN winds, mass conservation is ensured as follows: at each timestep, the mass accreted onto the black hole and injected as new wind particles (if applicable) are summed. The total mass is then removed using stochastic swallowing of gas particles within the black hole kernel (see \citealt{Angles-Alcazar} for more details).

\section{The dependence of physical galaxy properties on the AGN feedback model}\label{sec:FIRE_physical_properties}
\subsection{Evolution of stellar mass, gas mass and SFR}
In Figure \ref{fig:daniel_re_vs_t} (top panel), we plot
the evolution of stellar mass, star formation rate\footnote{Note that $\rm{SFR}_{10}$ is calculated using the total stellar mass formed in the last $10\,\rm{Myr}$ (and $\rm{SFR}_{100}$ from the mass formed within the last $100\,\rm{Myr}$), neglecting stellar mass loss for simplicity.} and gas mass (without cuts on phase or temperature), for the simulation without AGN winds and for simulations with five different wind strengths. These are calculated within a 3D aperture of radius $0.1\,R_{\rm{vir}}$. In all cases, the stellar mass increases monotonically, but the rate of increase is lower for higher feedback strengths. In the no-feedback case, the $10\,\rm{Myr}$-averaged star formation rate increases from $\sim70\,\rm{M_{\odot}}\,yr^{-1}$ to $\sim200\,\rm{M_{\odot}}\,yr^{-1}$ within $35\,\rm{Myr}$. The addition of AGN feedback suppresses this, with weak feedback (m1e5) resulting in a gradual increase to $\sim100\,\rm{M_{\odot}}\,yr^{-1}$ before a slight downturn at $25\,\rm{Myr}$ and the strongest feedback suppressing the increase in star formation from $\sim10\,\rm{Myr}$ and lowering it to $\sim40\,\rm{M_{\odot}}\,yr^{-1}$ after $35\,\rm{Myr}$. The m0.1e0.5 run represents a scenario where the feedback has little impact on any of the quantities plotted, while the m4e20 and m10e50 runs approach `saturation' of the effects of feedback. In all the simulations, the gas mass decreases over the $35\,\rm{Myr}$. In the simulation without AGN-driven winds, this is due to depletion by SF and expulsion by stellar winds. More rapid reduction in gas mass is seen for the stronger feedback simulations, where gas expulsion (due to the added winds) is more significant.

\subsection{Evolution of intrinsic galaxy sizes}
In Figure \ref{fig:daniel_re_vs_t} (bottom panel), we plot (from left to right), radii containing half the stellar mass, SFR, and gas mass versus time, for the six simulation variants described in Table \ref{Table:sims_params}. These are calculated directly from the stellar and gas particle data. Variations in the feedback model have a significant impact on the spatial extent of all three quantities. In the fiducial model (without AGN feedback), the half-mass radius decreases from $\sim1\,\rm{kpc}$ to $\sim0.5\,\rm{kpc}$ in $\sim35\,\rm{Myr}$. This coincides with an increase of $\sim300\,\rm{km\,s^{-1}}$ in the circular velocity (calculated within $1\,\rm{kpc}$; \citealt{Angles-Alcazar}). Increasing the kinetic feedback efficiency has a dramatic effect on the evolution of the half-mass radius. A value of $\epsilon_{k}=0.1$ (mass loading factor $\dot{M}_{\rm{w}}=44.4\,\rm{M_{\odot}\,yr^{-1}}$, simulation m2e10) is sufficient to maintain a roughly constant half-mass radius, while stronger feedback can slightly increase the size relative to the $t=0$ starting point. The radius containing half of the recently-formed stars, $R_{\rm{1/2},\,SFR_{10\,\rm{Myr}}}$ changes even more sharply, decreasing by a factor of $\sim4$ without AGN feedback. As for the half-mass radius, stabilisation is observed for the intermediate feedback (m2e10) run, with stronger feedback resulting in an increased $R_{\rm{1/2},\,SFR_{10\,\rm{Myr}}}$. The radius containing half of the gas mass,  $R_{\rm{1/2},\,\rm{gas}}$, is particularly sensitive to the feedback model. AGN feedback drives gas further from the center of the galaxy, resulting in $R_{\rm{1/2},\,\rm{gas}}\sim4-8\,\rm{kpc}$ after $\sim35\,\rm{Myr}$, compared to $R_{\rm{1/2},\,\rm{gas}}\sim0.2\,\rm{kpc}$ for the fiducial run.

\begin{figure*} 
\centering
\includegraphics[width=0.98\columnwidth]{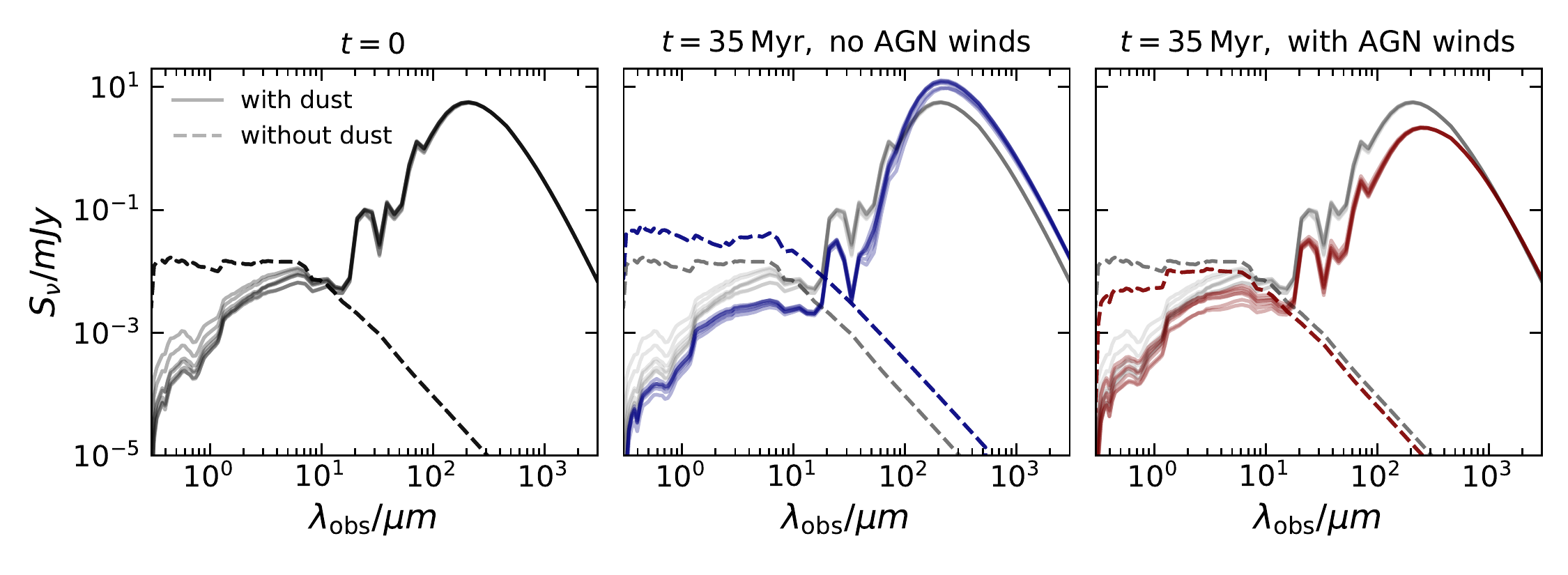}
\vspace{-0.4cm}
\caption{The impact of AGN winds on a galaxy's observed-frame SED. Left: the {\sc skirt}-predicted SED for the galaxy at $t=0$, before the AGN winds are turned on (black). Center: the predicted SED for the same galaxy $t=35\,\rm{Myr}$ later, for the case without AGN feedback (blue). At this time, the galaxy experiences a starburst, with boosted FIR emission and lower emission at shorter wavelengths, relative to $t=0$. Right: the predicted SED at $t=35\,\rm{Myr}$, for the case with AGN-driven winds (red). On all panels, solid lines show the dust-attenuated total emission and dashed lines show the intrinsic stellar emission, before dust attenuation. We overplot the $t=0$ SEDs on the two $t=35\,\rm{Myr}$ panels for easier comparison. The winds suppress the dusty starburst, lowering the overall star formation rate, so the predicted FIR emission is lower than at $t=0$. Different lines of the same colour represent different viewing angles.}
\label{fig:sed_0_35Myr}
\end{figure*}

\section{The impact of AGN-driven winds on the observed-frame UV-FIR continuum emission}\label{sec:results}
From here on, we focus on modelling and comparing the multi-wavelength emission from just two of the seven simulations: {\bf{noAGN}}, the fiducial model with no AGN-driven winds, and {\bf{m2e10}}, a model with intermediate-strength winds. As shown in Figure \ref{fig:daniel_re_vs_t}, this is the weakest wind implementation that maintains the half-mass radius (of both total stellar mass and mass of young stars) of the galaxy at an approximately constant value across the $35\,\rm{Myr}$ studied. In this model, a black hole of mass $10^{9}\,\rm{M_{\odot}}$ is placed at the center of the galaxy at $t=0$ (corresponding to $z=2.28$). For the subsequent $35\,\rm{Myr}$, the accretion rate is fixed at the Eddington accretion rate, $\dot{M}_{\rm{BH}}=22.2\,\rm{M_{\odot}\,yr^{-1}}$, corresponding to $L_{\rm{BH}}=1.26\times10^{47}\,\rm{erg\,s^{-1}}$. The kinetic feedback efficiency $\epsilon_{k}=0.1$. The AGN wind mass loading factor $\eta=2$, hence the mass outflow rate is $\dot{M}_{\rm{w}}=44.4\,\rm{M_{\odot}\,yr^{-1}}$. In our fiducial model without AGN-driven winds, the black hole still accretes at $\dot{M}_{\rm{BH}}=22.2\,\rm{M_{\odot}\,yr^{-1}}$, but there is no feedback.

\subsection{Synthetic observations of continuum emission}\label{sec:synthetic_obs}
We model the observed-frame spectral energy distributions (SEDs) at every snapshot, between $t=0\,\rm{Myr}$ and $t=35\,\rm{Myr}$, for the simulated galaxy with and without the AGN-driven wind. Following \cite{Cochrane2019,Cochrane2022a,Cochrane2022} and \cite{Parsotan2021}, we use the {\sc{skirt}}\footnote{\url{http://www.skirt.ugent.be}} radiative transfer code \citep{Baes2011,Camps2014} to make predictions for emission between rest-frame ultraviolet (UV) and far-infrared (FIR) wavelengths. For all snapshots, gas and star particles within $0.1R_{\rm{vir}}$ are drawn directly from the simulation. Dust particles are assumed to follow the distribution of the gas particles, with a dust-to-metals mass ratio of $0.4$ \citep{Dwek1998,James2002}. We assume dust destruction at $>10^6\,{\rm K}$ \citep{Draine1979,Tielens1994}. We model a mixture of graphite, silicate and PAH grains using the \cite{Weingartner2001} Milky Way dust prescription. Star particles are assigned \citet{Charlot2003} SEDs based on their ages and metallicities, assuming a \cite{Charlot2003} Initial Mass Function. We perform the radiative transfer on an octree dust grid, in which cell sizes are adjusted according to the dust density distribution, with the condition that no dust cell may contain more than $0.0001\%$ of the total dust mass of the galaxy. Dust self-absorption is modelled. The output from {\sc{skirt}} comprises predictions for global galaxy SEDs as well as maps of the resolved emission at each of the $\sim100$ wavelengths modelled. Convergence tests confirm that our various {\sc{skirt}} parameter choices (number of photon packages, maximum fraction of dust in one cell, minimum and maximum dust grid level) are sufficient \citep{Cochrane2022}.

\begin{figure*} 
\centering
\includegraphics[width=\columnwidth]{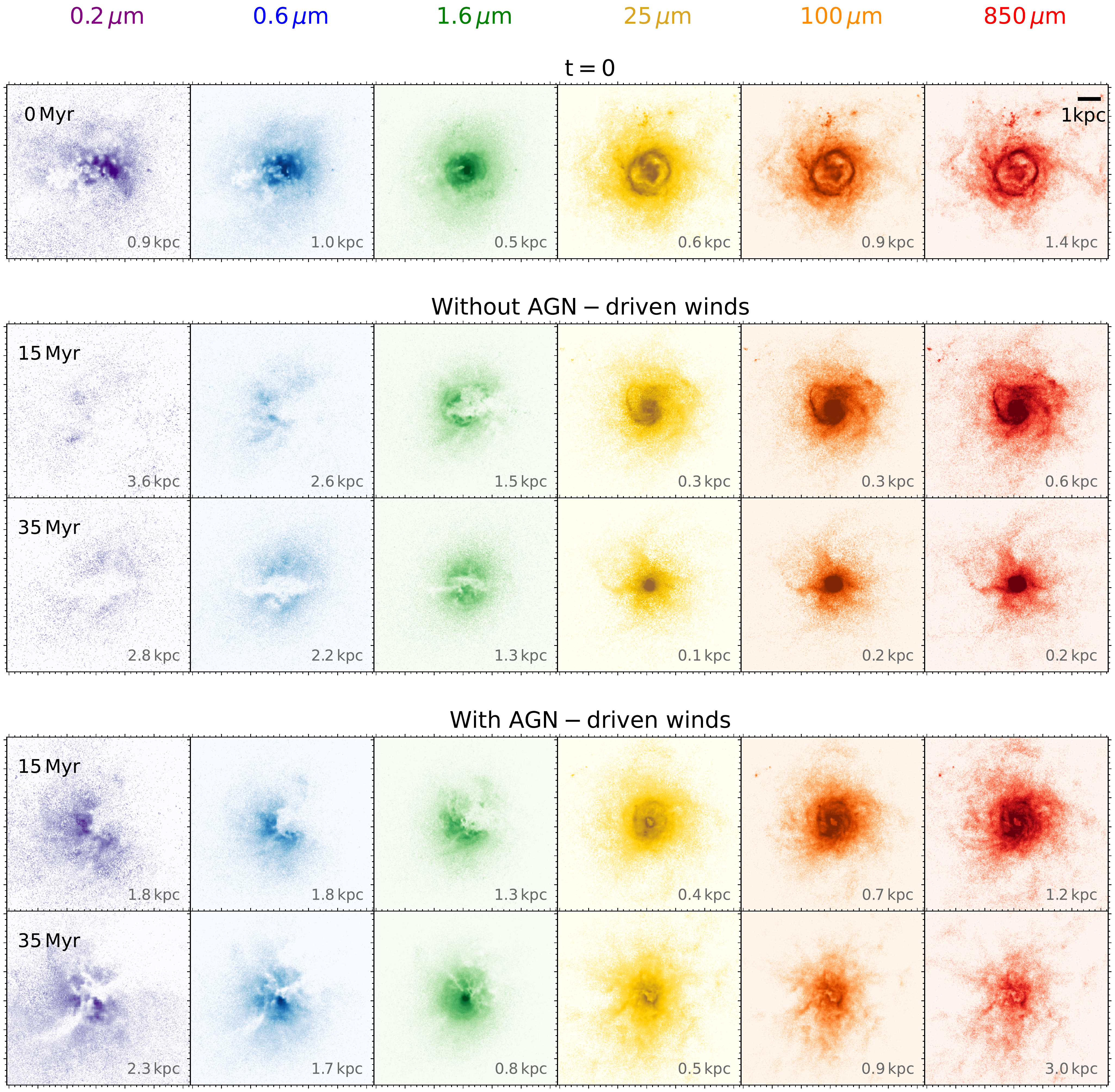}
\caption{The impact of AGN-driven winds on spatially-resolved, multi-wavelength emission. From left to right, we show {\sc skirt}-predicted rest-frame UV to rest-frame FIR emission. The top panel shows predicted emission at $t=0$, before the AGN winds are turned on. We also show predictions for the emission $t=15\,\rm{Myr}$ and $t=35\,\rm{Myr}$ later, for the same galaxy with and without AGN winds. In the absence of AGN-driven winds, the galaxy experiences a dusty starburst event, with extremely compact FIR emission at $t=35\,\rm{Myr}$. The AGN winds suppress the dusty starburst, leading to fainter FIR emission; the differences between the FIR emission in the two models are seen most clearly in the central regions of the galaxies. At shorter wavelengths, the galaxy appears brighter in the case with AGN winds at $35\,\rm{Myr}$, due to lower amounts of dust obscuration. Every panel in a given column is plotted on the same colour-scale, for easy comparison of relative fluxes. Half-light radii calculated using simple apertures are noted in grey.}
\label{fig:observable_maps}
\end{figure*}

\subsection{Predicted galaxy SEDs with and without AGN-driven winds}\label{sec:sed_results}
In Figure \ref{fig:sed_0_35Myr} we show {\sc{skirt}}-predicted galaxy SEDs for the fiducial simulated galaxy at $t=0$ (left-hand panel), for several viewing angles (black). We also show the SED for the two simulations at $t=35\,\rm{Myr}$, without AGN-driven winds (blue; central panel) and with AGN-driven winds (red; right-hand panel). On all three panels, dashed lines show the intrinsic (dust-free) SED. Without AGN winds, the {\it{intrinsic}} emission is boosted with respect to the $t=0$ SED, reflecting the increasing star formation rate over the $t=35\,\rm{Myr}$ (see Figure \ref{fig:daniel_re_vs_t}). however, the {\it{observable}} short wavelength emission ($\lambda_{\rm{rest}}\lesssim100\,\mu\rm{m}$) at $t=35\,\rm{Myr}$ is reduced with respect to the $t=0$ emission. This is due to dust attenuation, and in spite of the increased SFR. At longer wavelengths, the emission is boosted, reflecting a short period of intense, dusty starburst activity. \\
\indent When AGN-winds are implemented, the predicted intrinsic SED lies modestly below that at $t=0$, reflecting the suppression of the starburst. At all wavelengths, predicted observable emission is lower than at $t=0$. This is particularly notable at $\lambda_{\rm{rest}}\gtrsim100\,\mu\rm{m}$); FIR/sub-mm emission is reduced, and the peak of the FIR SED is also shifted longward. These two effects reflect the suppression of the dusty starburst, and associated colder dust temperatures. Interestingly, by $35\,\rm{Myr}$, the predicted {\it{observable}} short-wavelength emission is higher when winds are implemented than when they are not, despite the high SFR in the no-winds case driving higher {\it{intrinsic}} emission; again, this is due to heavy dust obscuration in the no-winds case.

\begin{figure*} 
\centering
\includegraphics[scale=0.65]{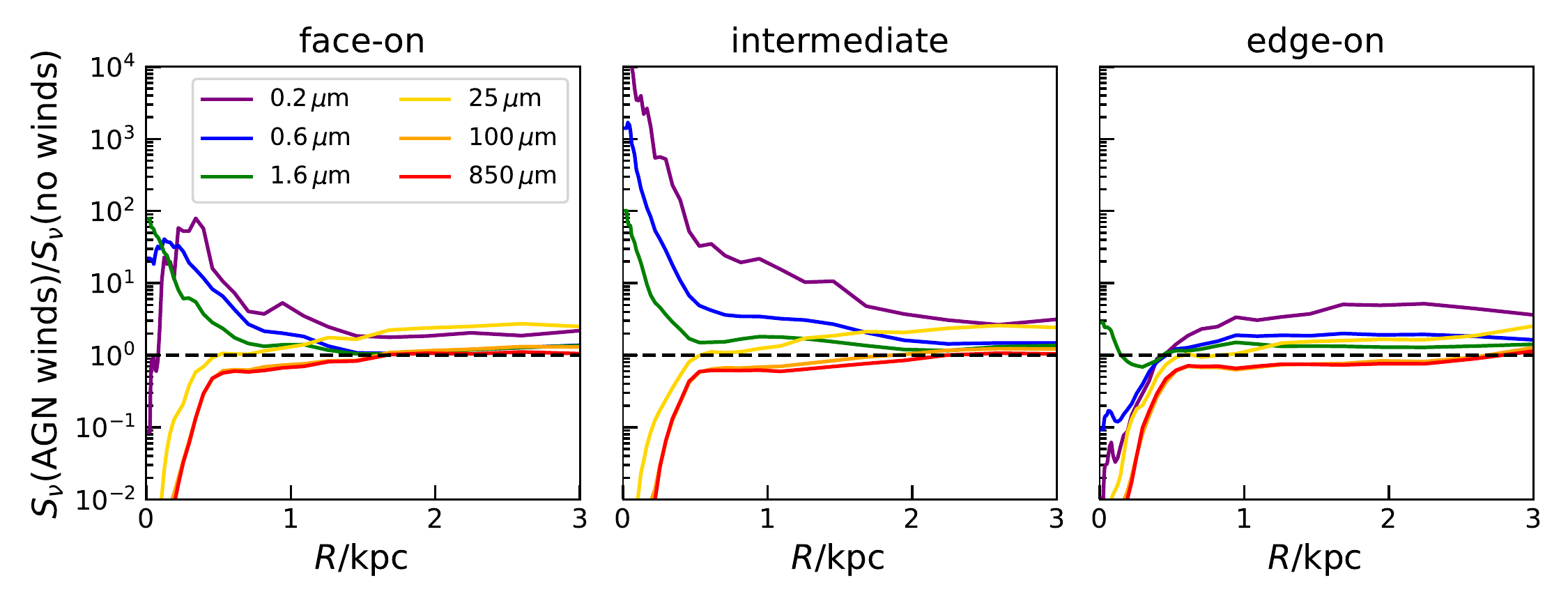}
\vspace{-0.3cm}
\caption{Ratios of predicted observable light within radial bins, for the case with AGN feedback, compared to the case without, $35\,\rm{Myr}$ after starting the two simulations. The three panels show three orientations of the galaxy with respect to the observer, roughly (from left-to-right) face-on (the orientation shown in Figure \ref{fig:observable_maps}), $\sim45\,\rm{deg}$ and edge-on. For the simulation with AGN winds included, the predicted flux at NIR and shorter wavelengths can be strongly enhanced relative to the no-feedback case, at radii $\lesssim1.5\,\rm{kpc}$, by up to factors of $10^{4}$. This is particularly interesting given that the stellar mass and SFR are both lower in the simulation with AGN winds; i.e. the simulation with {\it{higher}} stellar mass and SFR displays {\it{lower}} short-wavelength emission. We will study the role of dust in driving this effect in Figure \ref{fig:one_orientations_without_dust}. In contrast, at MIR and longer wavelengths, the flux is suppressed by AGN winds in the inner regions.}
\label{fig:three_orientations_profiles}

\end{figure*}
\begin{figure}
\centering
\includegraphics[width=0.93\columnwidth]{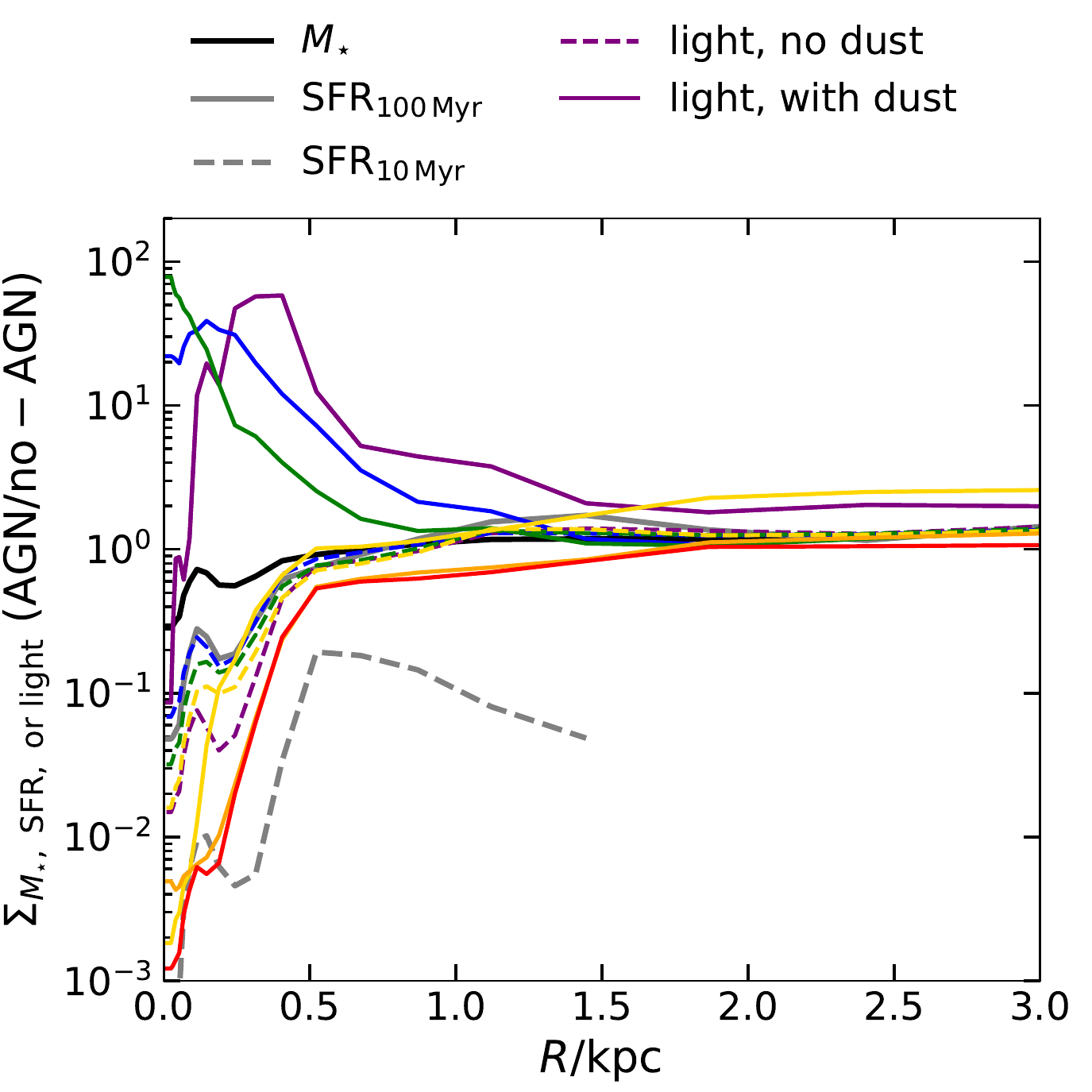}
\caption{The effects of dust attenuation on observed light profiles. The face-on panel from Figure \ref{fig:three_orientations_profiles} is reproduced here, with additional dashed coloured lines comparing the modelled intrinsic emission, before dust attenuation. Thicker black (and grey) lines show ratios of the projected face-on stellar mass (and both $100\,\rm{Myr}$-averaged and $10\,\rm{Myr}$-averaged star formation rate profiles) at $35\,\rm{Myr}$ in the case with AGN feedback, to the case without. The ratio of {\it{intrinsic}} short-wavelength emission broadly follows the ratio of the recent star formation: at $R\lesssim1\,\rm{kpc}$, the intrinsic emission in the case of AGN winds is {\it{lower}} than the case without winds, as the winds drive gas outwards, suppressing central star formation. However, the {\it{observable}} short-wavelength emission follows the opposite trend: the measured emission in the case of AGN winds is {\it{higher}} than in the case without winds, due to dust attenuation. This is because the short-wavelength light from the dusty starburst is much more strongly attenuated close to the center of the galaxy, for the case without AGN-driven winds.}
\label{fig:one_orientations_without_dust}
\end{figure}

\begin{figure*}
\centering
\includegraphics[width=\textwidth]{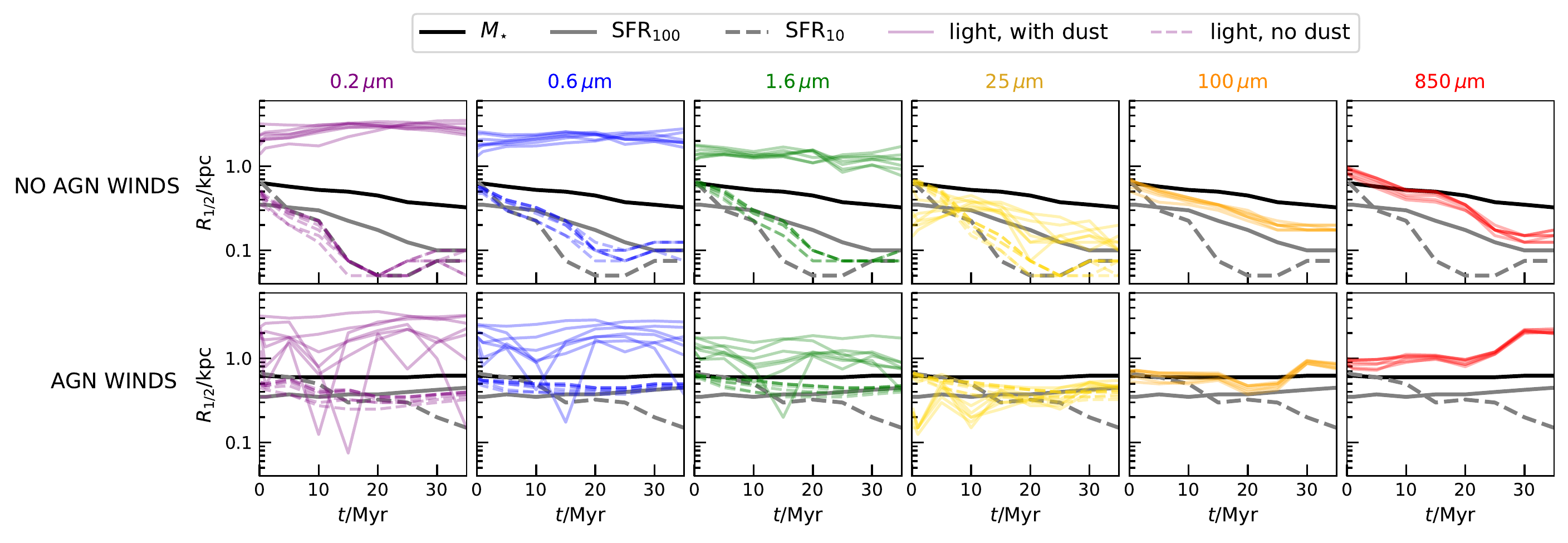}
\caption{Upper: Half-light radius versus time for six wavelengths shown on six panels, for the simulation without AGN feedback. Different coloured lines represent different viewing angles. We overplot the evolution of half-stellar mass (solid black lines) and half-SFR radii (solid and dashed grey lines show this calculated for SFR within the previous $100\,\rm{Myr}$ and $10\,\rm{Myr}$, respectively). The half-light radius of the rest-frame UV and optical observable emission remains approximately constant at $\sim2\,\rm{kpc}$, despite a rapid decrease in the extent of the stellar mass, the SFR, and thus the intrinsic emission (modelled {\it{without}} dust attenuation, as indicated by the coloured dashed lines). Lower: as the upper panel, but for the simulation with AGN wind-driven feedback. The winds maintain a roughly constant half-mass and half-$\rm{SFR}_{100\,\rm{Myr}}$  radius, which is reflected in the stable half-light sizes at all wavelengths. Short wavelength emission shows larger variation in measured size with orientation.}
\label{fig:size_vs_time_all}
\end{figure*}

\subsection{Spatially-resolved emission with and without AGN-driven winds}\label{sec:results_maps}
In Figure \ref{fig:observable_maps} we show predictions for the observed-frame optical (rest-frame UV) to observed-frame sub-mm (rest-frame FIR) emission for the galaxy at $t=0$ ($z=2.32$) in the first row. In the other panels, we show the predicted emission $t=15\,\rm{Myr}$ and $t=35\,\rm{Myr}$ later, for the simulation with and without AGN-driven winds. In all cases, the galaxy is aligned with respect to the gas angular momentum vector so that it is viewed from an approximately `face-on' orientation. \\
\indent In the simulation without AGN-driven winds, the observed-frame mid- and far-IR emission becomes progressively more compact, forming a core of bright emission by $t=35\,\rm{Myr}$. Spiral arms that are visible at $t=15\,\rm{Myr}$ have disappeared by $t=35\,\rm{Myr}$. At near-IR and shorter wavelengths, emission is strongly suppressed in the central regions, at both $t=15\,\rm{Myr}$ and $t=35\,\rm{Myr}$. Relative to $t=0$, the rest-frame UV/optical emission is  fainter (as seen in Figure \ref{fig:sed_0_35Myr}; in Section \ref{sec:results_profiles}, we will see that that this is driven by increased dust attenuation). Only faint emission in the outskirts is visible. This is in line with observations of dusty star-forming galaxies at similar redshifts \citep[e.g.][]{Hodge2016,Cochrane2021} as well as with our previous simulations \citep{Cochrane2019}. \\
\indent When AGN-driven winds are implemented, gas is evacuated from the center of the galaxy, generating a small central cavity by $t=15\,\rm{Myr}$ and suppressing the compact starburst \citep{Mercedes-Feliz2022a}. By $t=35\,\rm{Myr}$, the sub-mm emission actually appears more extended than at $t=0$, in contrast to the no-winds simulation. The changes between emission seen at $t=0$ and at $t=35\,\rm{Myr}$ are more modest at shorter wavelengths. In Section \ref{sec:results_profiles}, we quantify the impact of the winds on the radial profiles of light, from various viewing angles.

\subsection{Quantifying the impact of AGN-driven winds on radial profiles of light}\label{sec:results_profiles}
As seen in Section \ref{sec:results_maps}, the predicted maps of multi-wavelength emission as viewed from a `face-on' orientation differ substantially between our AGN winds and no-winds model. To quantify this further, and to look at the dependence of the measured emission on line-of-sight, we derive radial profiles of the predicted emission, as seen from three different viewing angles (approximately `face-on' at $0\,\rm{deg}$, $45\,\rm{deg}$, and approximately `edge-on' at $90\,\rm{deg}$). In Figure \ref{fig:three_orientations_profiles}, we show the ratio of the radial profiles for the simulations with AGN-driven winds to those without winds, both at $t=35\,\rm{Myr}$. The impact of AGN feedback is seen across the whole extent of the galaxy, and is particularly strong closest to the center. Interestingly, the predicted flux at near-IR and shorter wavelengths can be strongly enhanced relative to the no-feedback case, by up to factors of $10^{4}$. This is the case for most of the viewing angles simulated, but not all; the edge-on view shown in the right-hand panel actually shows suppression of short-wavelength emission in the inner regions. The differences between the profiles of short-wavelength emission with orientation hint at dust attenuation playing a key role. In the edge-on case, there is very little short-wavelength emission in either the winds or no-winds simulation; even in the case with AGN winds, the approximately edge-on disk is sufficiently optically thick that little short-wavelength light escapes.\\
\indent In Figure \ref{fig:one_orientations_without_dust}, we illustrate more directly the role of dust in driving the high ratios at $R\lesssim1\,\rm{kpc}$. We again plot the ratios between the $0.6-850\,\mu\rm{m}$ emission predicted by the simulations with AGN winds and those without, for the face-on orientation shown in Figure \ref{fig:three_orientations_profiles}. We additionally calculate the intrinsic `transparent' emission ({\it{without}} dust attenuation or emission). We calculate and plot the ratios between the emission predicted by the simulations with and without AGN winds using this intrinsic emission, for rest-frame wavelengths up to and including $25\,\mu\rm{m}$. We can thus pick apart the roles of differing intrinsic emission and differing dust attenuation. The ratios of stellar mass and recent star formation rate (calculated over the last $10\,\rm{Myr}$ and $100\,\rm{Myr}$) profiles are also plotted. In line with expectations, the ratios of the intrinsic emission approximately follow the ratio of the SFR. For the simulation with AGN feedback, SFR is strongly suppressed at $R\lesssim0.5\,\rm{kpc}$, relative to the no-feedback case, due to the winds driving the evacuation of gas from the innermost regions of the galaxy (see \citealt{Mercedes-Feliz2022a} for more detailed characterisation of the cavity generated by different AGN wind models). The {\it{observable}} short-wavelength emission follows the opposite trend: the measured emission in the case of AGN winds is {\it{higher}} than in the case without winds, because of dust attenuation. For the no-winds simulation, the short wavelength light is much more strongly attenuated than in the simulation with AGN winds, as the winds remove the dust from the central regions of the galaxy.

\begin{figure*} 
\centering
\includegraphics[scale=0.55]{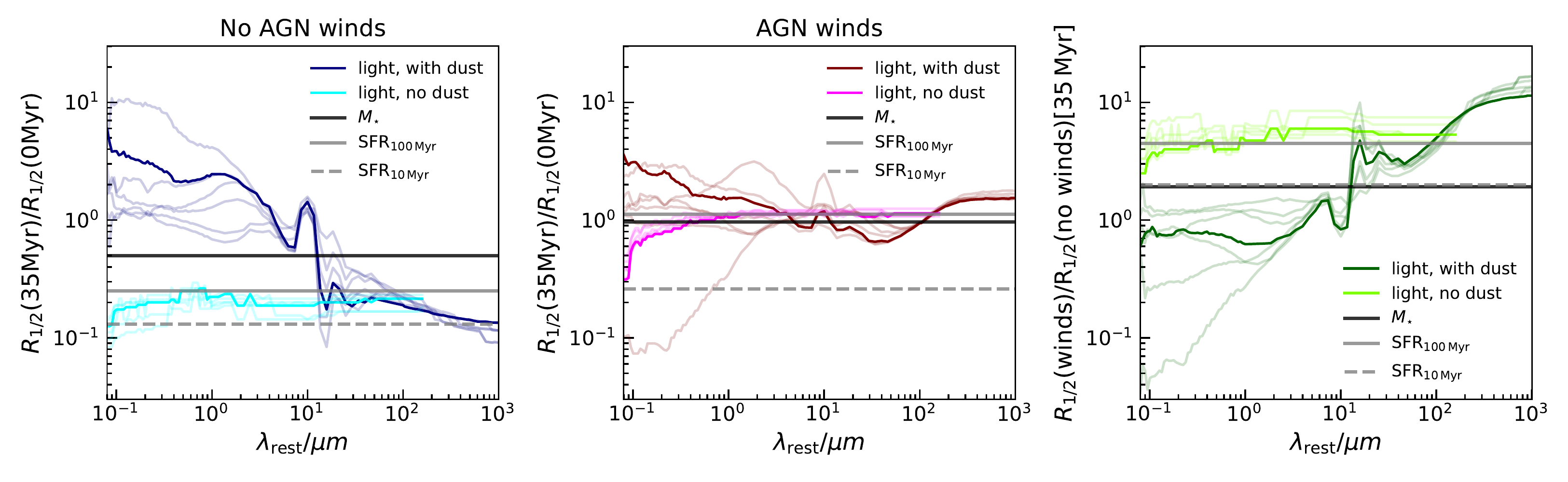}
\vspace{-0.5cm}
\caption{The ratio of the half-light radius (before and after dust attenuation) at $t=35\,\rm{Myr}$ to that at $t=0\,\rm{Myr}$, as a function of rest-frame wavelength, for the simulation without AGN-driven winds (left) and with AGN-driven winds (center), for different viewing angles (pale lines). In both panels, the black line shows the ratio of the half-mass radius at $35\,\rm{Myr}$ to that at $0\,\rm{Myr}$. The solid (dashed) grey horizontal line shows the ratio of the half-$\rm{SFR}_{100\,\rm{Myr}}$ (half-$\rm{SFR}_{10\,\rm{Myr}}$) radius at $35\,\rm{Myr}$ to that at $0\,\rm{Myr}$. For both the no-winds and winds simulations, the ratio of {\it{intrinsic}} emission (light emitted short-ward of the mid-IR, before dust attenuation) at $t=35\,\rm{Myr}$ to that at $t=0\,\rm{Myr}$ reflects the change in the half-$\rm{SFR}_{100\,\rm{Myr}}$ radius. For the simulation without AGN-driven winds, the change in half-light radius of {\it{observed}} emission (after dust attenuation) is more substantial: strong compact, dusty star formation results in compact FIR emission and extended faint UV-NIR emission. When AGN winds are implemented, gas is prevented from accumulating in the galaxy's center, and the half-light radii remain more stable over the $35\,\rm{Myr}$. Right: The ratio of the half-light radius measured at $t=35\,\rm{Myr}$ in the simulation with AGN feedback to that in the simulation without, as a function of rest-frame wavelength, for intrinsic and dust-attenuated emission. After $35\,\rm{Myr}$, AGN feedback boosts the half-light radius of intrinsic stellar emission relative to the case without feedback by a factor of $\sim4$, roughly following the boost in the half-$\rm{SFR}_{100\,\rm{Myr}}$  radius. Once dust attenuation is accounted for, the ratio of half-light radius in the AGN wind simulation to that of the no-wind simulation is strongly dependent on wavelength and viewing angle. The changes in radii at short wavelengths are line-of-sight dependent but tend to be more modest than at long wavelengths; there is a very clear trend of winds driving larger radii in the FIR.}
\label{fig:size_vs_wavelength_ratio}
\end{figure*}

\subsection{The impact of AGN-driven winds on half-light sizes}\label{sec:results_sizes}
Changes in the radial profiles of emission translate to changes in measured half-light radii. In this section, we compare the change in half-light radius between $t=0$ and $t=35\,\rm{Myr}$ for the simulations with and without AGN-driven winds. We define half-light radius here as the radius within which half of the total light (defined using an aperture of radius $5\,\rm{kpc}$) is contained. Note that this is different to fitting a parametric form to the emission, as was done in \cite{Parsotan2021} and for the PSF-convolved images described in Section \ref{sec:results_size_mass} of this paper. When studying the emission at very high spatial resolution, a simple half-light radius provides a better measure of spatial extent due to clumpy and irregular emission (as is widely seen in Figure \ref{fig:observable_maps}). \\
\indent In Figure \ref{fig:size_vs_time_all} (upper panel), we show half-light radius (for emission before and after dust attenuation) versus time for the six representative rest-frame wavelengths, for the simulation without AGN winds. The spatial extent of the rest-frame UV and optical observable emission (i.e. the emission modelled {\it{with}} dust attenuation) remains approximately constant between $t=0$ and $t=35\,\rm{Myr}$. In contrast, the extent of the intrinsic emission (i.e. the emission modelled {\it{without}} dust attenuation) decreases rapidly from $\sim0.4\,\rm{kpc}$ to $\sim0.05\,\rm{kpc}$ within $\sim20\,\rm{Myr}$, tracing the rapid decrease in the extent of star formation. At longer wavelengths, we see gentler compaction, presumably because emission traces longer timescales. In Figure \ref{fig:size_vs_time_all} (lower panel) we plot the same quantities, for the simulation with AGN-driven winds. At all wavelengths, the half-light radii of intrinsic and observable emission remain roughly constant, in line with the approximate constancy of the half-SFR radius. At UV and optical wavelengths, we see large variations in the derived half-light radius from one timestep to the next; this is due to the disturbed and
variable dust geometry.\\
\indent In Figure \ref{fig:size_vs_wavelength_ratio}, we plot the ratio of the half-light radius at $t=35\,\rm{Myr}$ to that at $t=0\,\rm{Myr}$ as a function of wavelength, for simulations without (left-hand panel) and with (center panel) AGN-driven winds. We show ratios of intrinsic (dust-free) emission and observable emission. We also show the ratio of three key physical properties (half-mass, half-$\rm{SFR}_{10\,\rm{Myr}}$ and half-$\rm{SFR}_{100\,\rm{Myr}}$  radii) at $35\,\rm{Myr}$ to that at $0\,\rm{Myr}$. For both simulations with and without AGN-driven winds, the ratio of {\it{intrinsic}} emission (light emitted short-ward of the mid-IR, before dust attenuation) at $35\,\rm{Myr}$ to that at $0\,\rm{Myr}$ closely traces the change in the half-SFR radius. For the simulation without AGN-driven winds, the change in half-light radius of {\it{observed}} emission (after dust attenuation) is more substantial. In this case, emission at $\lambda_{\rm{rest}}\gtrsim100\,\mu\rm{m}$ becomes a factor of $\sim10$ more compact within $35\,\rm{Myr}$, reflecting the increasingly compact starburst. However, strong attenuation causes an increase in the half-light radius of the observed short-wavelength emission in some cases, by up to a factor of $\sim10$. When AGN winds are implemented, gas is prevented from accumulating in the galaxy's center, and the half-light radii remain more stable over the $35\,\rm{Myr}$.\\
\indent In the right-hand panel of Figure \ref{fig:size_vs_wavelength_ratio}, we take the two models at $35\,\rm{Myr}$ (with and without AGN feedback), and calculate the ratio of the half-light radii as a function of wavelength. The half-light radius of the intrinsic stellar emission is boosted when the AGN-driven wind model is implemented, relative to the case without feedback. As discussed, this reflects star formation being distributed over a larger extent of the galaxy, since winds drive the gas outward (see also \citealt{Mercedes-Feliz2022a}); in the no-feedback case, emission is concentrated in the center. Without dust attenuation, there is little dependence of derived half-light radius on viewing angle at any wavelength. Once dust attenuation is taken into account, however, the ratio of half-light radius in the AGN wind simulation to that of the no-wind simulation is strongly dependent on wavelength and viewing angle. Importantly, after $35\,\rm{Myr}$, the rest-frame UV-NIR emission is more compact in the case with AGN winds; this is at odds with the increase in half-mass radius and non-intuitive. 

\subsection{Placing the simulated galaxies on the size-stellar mass relation}\label{sec:results_size_mass}
\begin{figure} 
\centering
\includegraphics[width=\columnwidth]{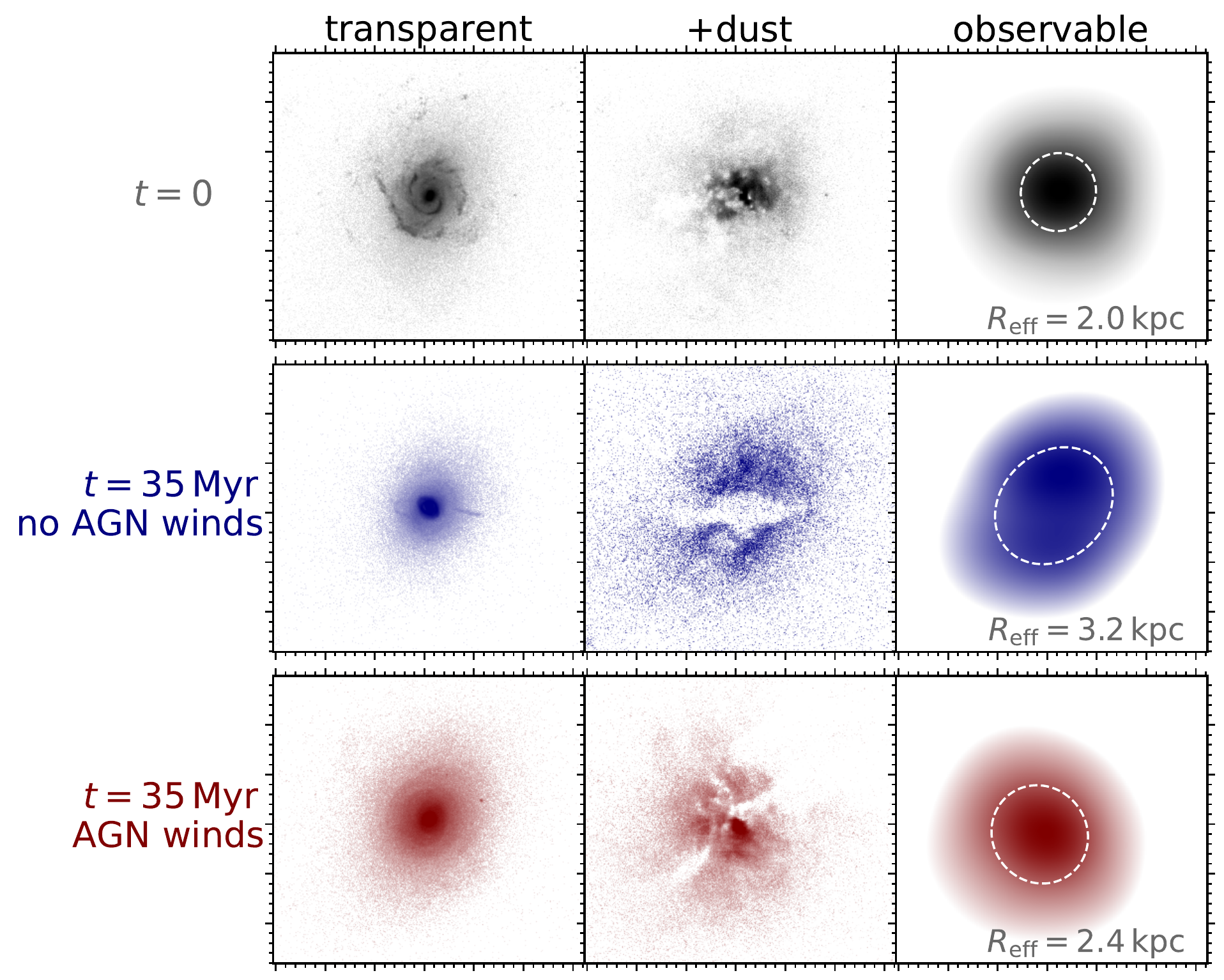}
\caption{Schematic showing dust-free, dust-attenuated and convolved rest-frame $B$-band images for the face-on no-wind simulation at $t=0$ (upper panel) and $t=35\,\rm{Myr}$ (center) and for the AGN wind simulation at $t=35\,\rm{Myr}$. Sizes stated on the right-hand panels are the semi-major axes inferred from S\'ersic profile fits to the convolved images (highlighted with white dashed lines).}
\label{fig:image_convolution}
\end{figure}

We have seen that the addition of AGN winds to the FIRE simulations can suppress a period of intense compact star formation and associated reduction in physical galaxy size. We have also seen that the observable manifestations of this are complex: because of dust attenuation, the half-light radii at UV-NIR wavelengths are actually lower for the AGN wind simulation, once dust is modelled. In this section, we study the impact of these effects on the inferred locations of galaxies on the size-stellar mass relation. \\
\indent Following \cite{Parsotan2021}, we derive effective radii comparable to those used to infer the size-mass relation from observations. We take the predicted rest-frame $5000\,\angstrom$ emission maps (before and after dust attenuation), and convolve them with the HST WFC3/IR F160W point spread function (\citealt{Skelton2014}; F160W was chosen to match the observed-frame wavelength of the rest-frame $5000\,\angstrom$ emission at $z=2.3$). We resample each convolved image onto a $0.06''$ pixel scale (see Figure \ref{fig:image_convolution}) and insert the resulting image into a blank region of a randomly-selected CANDELS HST F160W image. This produces synthetic HST images with realistic correlated noise. We then fit these images with a 2D S\'ersic profile. The effective radii ($R_{e}$) derived are the semi-major axes of the fitted S\'ersic ellipses, as in the observational work we compare to \citep[see][]{VanDerWel2014}. \\
\begin{figure*} 
\includegraphics[width=1.0\textwidth]{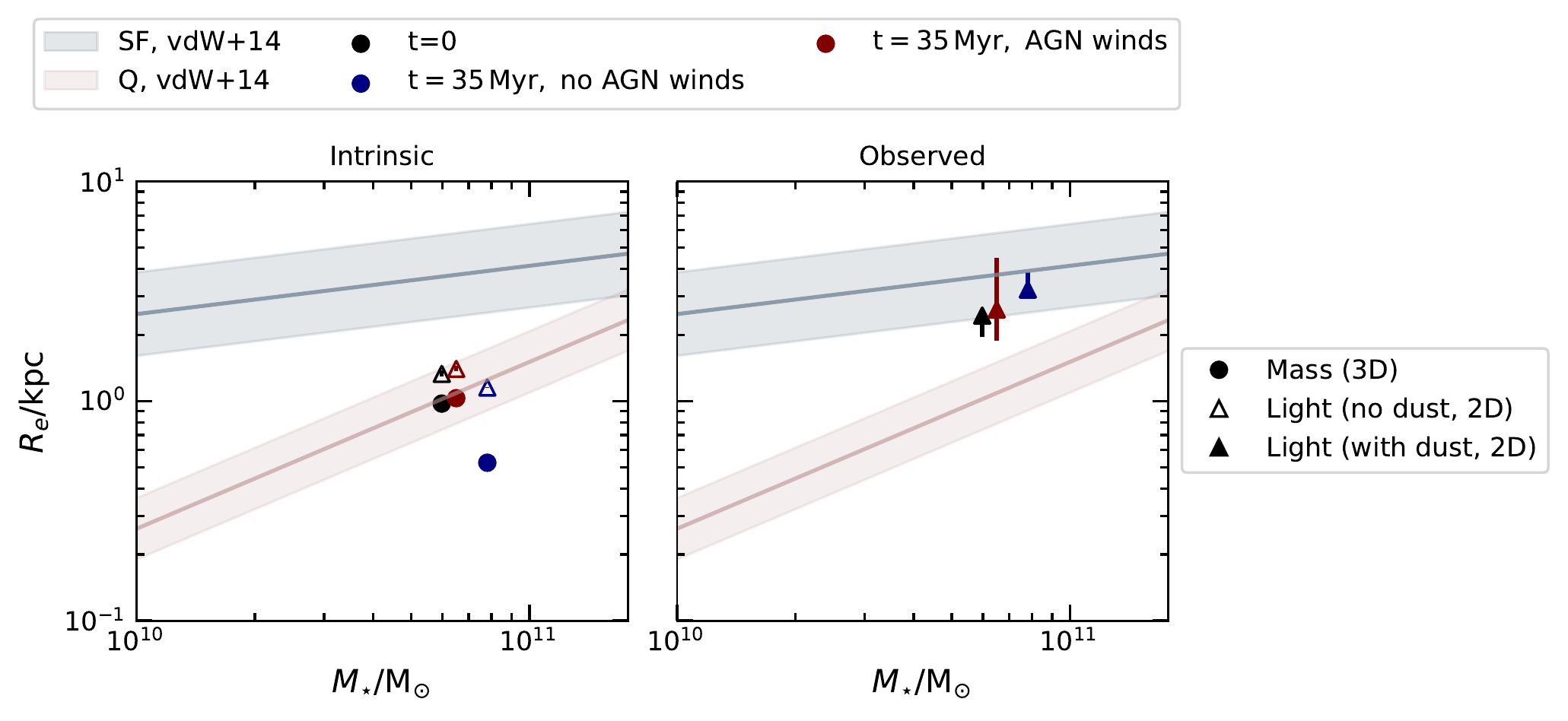}
\caption{The stellar masses and sizes of our simulated galaxy at $t=0$ (black) and at $t=35\,\rm{Myr}$ for the no wind (blue) and AGN wind (red) simulations, derived using several different methods, with observationally-inferred relations for star-forming and quiescent galaxies at $z\sim2.25$ \citep{VanDerWel2014} overlaid. Left: Filled circles show half-mass sizes, derived from the 3D stellar particle data. We show the effective radii of the {\it{intrinsic emitted}} $5000\,\angstrom$ light (without dust, the `transparent’ case) in open triangles. Right: Filled triangles show the effective radii that would be inferred with the effects of dust included. Error bars show variance in the size inferred from different orientations. At $t=0$, the observable effective radius at $5000\,\angstrom$ (including modelling dust) is $\sim2\,\rm{kpc}$, more than twice the half-mass radius (black symbols). By $t=35\,\rm{Myr}$, the half-mass radius has approximately halved to $\sim0.5\,\rm{kpc}$ for the case without AGN feedback (blue symbols). However, the extent of the {\it observable} emission at $5000\,\angstrom$ (blue triangle) is increased with respect to $t=0$, due to preferential attenuation towards the centre. When AGN emission is modelled (red symbols), the intrinsic and inferred radii change little between $t=0$ and $t=35\,\rm{Myr}$. The large (up to order of magnitude) differences between intrinsic and observable sizes (primarily due to the effects of dust during the short-lived dusty starburst phase) highlight the critical need to forward-model observables, including modelling dust attenuation, when studying the impact of variations in subgrid models on galaxy sizes.}
\label{fig:size-mass-relation}
\end{figure*}
\indent In the right-hand panel of Figure \ref{fig:size-mass-relation}, we show effective radii derived at $t=0$ and at $t=35\,\rm{Myr}$ for the model with and without AGN-driven winds, alongside the observationally-derived size-mass relations for star-forming and quiescent sources \citep{VanDerWel2014}. We label this panel, where radii are derived from S\'ersic profile fits to the convolved and re-projected $5000\,\angstrom$ image, generated including the effects of dust, as `observed'. In the left-hand panel, we plot half-mass radii calculated from the 3D particle data, alongside the effective radius derived from the $5000\,\angstrom$ image, generated from the `transparent'
 image {\it{without}} dust included. \\
\indent At $t=0$ (black symbols), the observable effective radius at $5000\,\angstrom$ is $\sim2\,\rm{kpc}$, $\sim2.5\,\times$ the half-mass radius. When the simulation is run without AGN-driven winds (blue symbols), the half-mass radius approximately halves to $\sim0.4\,\rm{kpc}$ in only $\sim35\,\rm{Myr}$. However, the extent of the {\it{observable}} emission at $5000\,\angstrom$ is increased with respect to $t=0$, due to preferential attenuation towards the centre. The rapid decrease in the half-mass radius seen in the dusty starburst is actually not reflected in the effective radius inferred from $5000\,\angstrom$ emission. When AGN winds are included (red symbols), the half-mass radius and effective radius of the $5000\,\angstrom$ emission change little between $t=0$ and $t=35\,\rm{Myr}$. As seen in Figure \ref{fig:size_vs_wavelength_ratio}, {\it{far-infrared wavelengths provide a better observable discriminant between the wind and no-wind models}}.\\
\indent The discrepancy between observed and intrinsic sizes is greatest in the no-winds case, which is experiencing a highly-obscured period of rapid star formation. We would expect this discrepancy to diminish following the dusty starburst event. While it would be infeasible to run our hyper-refined simulations over cosmological timescales to study this, we can glean some insights from the fiducial simulation, which was run until $z\sim1$ without AGN winds. As seen in Figure \ref{fig:fiducial_fig}, following a rapid decrease in half-mass radius to $\sim0.3\,\rm{kpc}$, the galaxy appears to stabilise with a half-mass radius of $\sim1\,\rm{kpc}$. We re-run our synthetic observation pipeline on the snapshot at $z=1.5$. At this redshift, the stellar mass is $\log_{10}(M_{\star}/\rm{M_{\odot}})=11.28$ and the half-mass radius is $1.02\,\rm{kpc}$. The effective radius of the $5000\,\angstrom$ emission without dust is $\sim1.3\,\rm{kpc}$. With dust included, the inferred effective radius is $\sim2.6\,\rm{kpc}$. While still a factor of $\sim2.5$ difference, this is not as extreme as the factor of $\sim5$ difference between half-mass size and measured effective radius seen during the dusty starburst phase.\\
\indent One important result of this work is that galaxies can display similar half-light radii at rest-frame optical wavelengths despite having very different half-mass sizes. This highlights the complex relationship between mass and light, and the difficulties inherent to observational studies of galaxy sizes. Spatially-resolved SED fitting \citep[e.g.][]{Suess2019,Abdurrouf2021a,Abdurrouf2022,Abdurrouf2022a} may present a promising path towards more robust estimates of half-mass sizes.\\

\section{Discussion and conclusions}\label{sec:conclusions}
We have presented a study of the effects of quasar-driven winds on the physical size of and observable emission from a massive, star-forming galaxy at $z=2.3$. This redshift was identified as the time at which the galaxy undergoes a very strong starburst and rapid decrease in half-mass size in the fiducial FIRE simulation, which does not model AGN feedback. Our new simulations include a novel hyper-refinement scheme that enables us to track the propagation of winds from scales of the inner nuclear region to the CGM and to characterise their impact on the multi-phase ISM modelled within the FIRE simulations. The driven wind is constant over the short period of time simulated (as opposed to being self-consistently tied to the time-varying black-hole accretion rate). This paper is part of a series analysing these simulations with AGN winds \citep[see also][]{Angles-Alcazar,Mercedes-Feliz2022a,Mercedes-Feliz2022,Terrazas2023}. Our key results are summarised here.\\
\indent We find that AGN winds are capable of evacuating gas from the inner regions of the galaxy and suppressing the compact, dusty starburst observed in the fiducial, no-feedback simulation. While previous models have suggested that AGN feedback can positively trigger starburst activity in galaxies \citep[e.g.][]{Ciotti2007,Bieri2015,Bieri2016}, the effect seen in our simulations is primarily negative \citep[see][for further discussion]{Mercedes-Feliz2022a}. Changes in the feedback model manifest as substantial differences in galaxy half-mass radius. In the fiducial model without AGN winds, the half-mass radius decreases from $\sim1\,\rm{kpc}$ to $\sim0.5\,\rm{kpc}$ within just $35\,\rm{Myr}$. AGN-driven wind implementations with different kinetic feedback efficiencies result in very different half-mass radii, with the strongest winds resulting in a $>10\%$ increase in half-mass radius within $35\,\rm{Myr}$. These results are qualitatively in line with earlier simulations \citep[e.g.][]{Dubois2013,Dubois2016,Choi2018}. We focus on simulations implementing an intermediate strength wind, with mass outflow rate $44.4\,\rm{M_{\odot}\,yr^{-1}}$ and kinetic feedback efficiency $0.1$; in this case, the galaxy maintains an approximately constant half-mass and half-SFR radius across the $\sim35\,\rm{Myr}$ studied; as a result, the intrinsic (dust-free) emission of a simulated galaxy without AGN winds is $\sim5\,\times$ more compact. \\
\indent Since our high-resolution simulations resolve the ISM in detail, we are able to perform careful modelling of radiative transfer effects. We model the rest-frame UV-to-FIR emission from both the fiducial and AGN winds simulations at various timesteps between $t=0$ and $t=35\,\rm{Myr}$ using the {\sc{skirt}} radiative transfer code. The SEDs of the simulated galaxy with and without AGN-driven winds display notable differences. At $35\,\rm{Myr}$, the galaxy without AGN winds displays moderately bright sub-millimeter emission ($S_{\nu}=1.1\,\rm{mJy}$ at $\lambda_{\rm{obs}}=870\,\mu\rm{m}$). With AGN winds turned on, $S_{\nu}=0.4\,\rm{mJy}$ at $\lambda_{\rm{obs}}=870\,\mu\rm{m}$, comparable to the brightness at $t=0$. In both the simulations with and without AGN-driven winds, UV-mid-IR flux is suppressed with respect to $t=0$ by a factor of a few. For the simulation without AGN winds, this happens in spite of the high SFR, and is due to substantial dust attenuation. When AGN winds are switched on, this happens because of the suppression of star formation, as gas is evacuated from the galaxy.\\
\indent The predicted emission maps for the simulations with and without AGN-driven winds display striking visual differences. In the model without AGN winds, the galaxy rapidly becomes more compact in the far-infrared, shrinking to $\sim0.2\,\rm{kpc}$. This contrasts starkly with the diffuse, extended rest-frame UV and optical emission, which is biased to larger radii after $35\,\rm{Myr}$ due to preferential dust attenuation in the galaxy's center. This is in qualitative agreement with multi-wavelength observations of some dusty high-redshift sub-millimeter galaxies. When AGN winds are switched on, the UV-NIR emission is boosted at radii $\lesssim1\,\rm{kpc}$ compared to the no-winds case by $\sim35\,\rm{Myr}$. The half-light radius remains broadly constant with time at all wavelengths. \\
\indent Finally, we place our simulated galaxies on the size-stellar mass plane, alongside observationally-derived relations, by convolving the predicted rest-frame $5000\,\angstrom$ emission with an {\it{HST}} PSF. The half-mass sizes derived directly from simulated stellar particles lie well below the observationally-derived relation for star-forming galaxies at this redshift, with the $\sim35\,\rm{Myr}$ no-winds snapshot the most discrepant. However, the modelled effective radii at $t=0$ and $t=35\,\rm{Myr}$, for both the winds and no-winds model, are in good agreement with the size-mass relation of observed sources at the same redshift. This demonstrates that forward-modelling simulations is crucial to make robust comparisons with observational data. We particularly urge caution for studies focused on dusty star-forming galaxies, where observable short-wavelength size is highly biased.\\
\indent We have restricted our study to an investigation of the impact of AGN-driven winds on the physical properties and observable stellar emission within a short period of time in the evolution of a massive galaxy simulated with FIRE physics. This has enabled us to perform a detailed investigation into the non-intuitive effects of winds on the stellar continuum emission. Future work should address in more detail whether AGN winds can efficiently regulate galaxy sizes over longer periods. The recent study of FIRE-2 simulations including a multi-channel AGN feedback model, self-consistently tied to the black hole accretion rate starting from cosmological initial conditions, suggests that AGN feedback can indeed prevent the formation of extremely dense stellar cores (\citealt[]{Wellons2022}; see also Byrne et al., in prep.). It will be important for future work on this to include radiative transfer effects on observed sizes and to further explore how the results depend on details of the assumed black hole physics. Emission from the AGN itself will also impact continuum observations \citep[e.g.][]{McKinney2021} and extensions to this work could consider the AGN radiation source alongside the stellar radiation.

\section*{Acknowledgements}
We thank the anonymous referee for helpful suggestions. RKC, DAA and CCH are grateful for funding from the Flatiron Institute. The Flatiron Institute is supported by the Simons Foundation. DAA acknowledges support by NSF grants AST-2009687 and AST-2108944, CXO grant TM2-23006X, Simons Foundation Award CCA-1018464, and Cottrell Scholar Award CS-CSA-2023-028 by the Research Corporation for Science Advancement. CAFG was supported by NSF through grants AST-1715216, AST-2108230, and CAREER award AST-1652522; by NASA through grants 17-ATP17-0067 and 21-ATP21-0036; by STScI through grants HST-AR-16124.001-A and HST-GO-16730.016-A; by CXO through grant TM2-23005X; and by the Research Corporation for Science Advancement through a Cottrell Scholar Award. Support for PFH was provided by NSF Research Grants 1911233, 20009234, 2108318, NSF CAREER grant 1455342, NASA grants 80NSSC18K0562, HST-AR-15800. Numerical calculations were run on the Caltech compute cluster ``Wheeler'', allocations AST21010 and AST20016 supported by the NSF and TACC, and NASA HEC SMD-16-7592. JM is funded by the Hirsch Foundation.\\
\indent Some of the simulations presented in this work were run on the Flatiron Institute’s research computing facilities (Gordon-Simons, Popeye, and Iron compute clusters), supported by the Simons Foundation. Other simulations were run using Northwestern University’s compute cluster `Quest' and the Extreme Science and Engineering Discovery Environment (XSEDE), which is supported by NSF grant ACI-1053575.\\
\indent This research has made use of the SVO Filter Profile Service, supported from the Spanish MINECO through grant AyA2014-55216.

\section*{Data availability}
The data underlying this article will be shared on reasonable request to the corresponding author. FIRE-2 simulations are publicly available \citep{Wetzel2022} at \url{http://flathub.flatironinstitute.org/fire}.
Additional FIRE simulation data is available at \url{https://fire.northwestern.edu/data}. A public version of the \textsc{Gizmo} code is available at \url{http://www.tapir.caltech.edu/~phopkins/Site/GIZMO.html}.

\bibliographystyle{mnras}
\bibliography{Edinburgh}


\bsp	
\label{lastpage}
\end{document}